%% file: main.tex
\documentclass[conference]{IEEEtran}
\usepackage[square,sort,comma,numbers]{natbib}
\usepackage{hyperref}
\usepackage{multirow}
\usepackage{color}
\usepackage{multirow}
\usepackage{url}
\usepackage{balance}
\usepackage{fancyvrb}
\usepackage{graphicx} 
\usepackage{subcaption}
\usepackage{listings}
\usepackage{xcolor}
\usepackage{comment}
\usepackage[bottom]{footmisc}

\IEEEoverridecommandlockouts
\begin{document}

\title{Characterizing the Performance of Emerging Deep Learning, Graph, and High Performance Computing Workloads Under Interference}

\author{\IEEEauthorblockN{Hao Xu} 
\IEEEauthorblockA{College of William and Mary\\
hxu07@email.wm.edu}
\and
\IEEEauthorblockN{Shuang Song}
\IEEEauthorblockA{The University of Texas at Austin\\
songshuang1990@utexas.edu}
\and
\IEEEauthorblockN{Ze Mao}
\IEEEauthorblockA{University of Southern California\\
zemao@usc.edu}

}

\maketitle
\thispagestyle{empty}
\pagestyle{plain}

\begin{abstract}
\input{abstract}
\end{abstract}

\section{Introduction}
\input{introduction}
\label{sec:intro}

\section{Related Work}
\input{related}
\label{sec:related}

\section{Experimental Methodology}
\input{experiment}

\label{sec:exp}

\section{Running without Interference}
\input{run_no_interference}

\label{sec:nointerference}

\section{Running with Interference}
\input{run_interference}

\label{sec:interference}

\section{Provenance of Interference}
\input{interference_analysis}

\label{sec:analysis}

\section{Conclusion}
\input{conclusion}
\label{sec:conclusion}

\scriptsize
\bibliographystyle{ieeetr}
\bibliography{ref}

\end{document}

%% file: abstract.tex
Throughput-oriented computing via co-running multiple applications in the same machine has been widely adopted to achieve high hardware utilization and energy saving on modern supercomputers and data centers. However, efficiently co-running applications raises new design challenges, mainly because applications with diverse requirements can stress out shared hardware resources (IO, Network and Cache) at various levels. The disparities in resource usage can result in interference, which in turn can lead to unpredictable co-running behaviors.
To better understand application interference, prior work provided detailed execution characterization. However, these characterization approaches either emphasize on traditional benchmarks or fall into a single application domain.
To address this issue, we study 25 up-to-date applications and benchmarks from various application domains and form 625 consolidation pairs to thoroughly analyze the execution interference caused by application co-running. Moreover, we leverage mini-benchmarks and real applications to pinpoint the provenance of co-running interference in both hardware and software aspects. 

%% file: introduction.tex
Modern computer systems have evolved to employ powerful parallel architectures, including multi-core processors, multi-socket chips, large memory subsystems, and fast network communication. Given such powerful hardware, it is difficult for a single application to achieve the theoretical peak performance. For example, many high-performance computing (HPC) software applications achieve only 5-15\% of the peak performance on modern supercomputers~\cite{lowutilization1}. 
To maximize the hardware usage, researchers consolidate multiple programs onto a single machine, which is known as throughput-oriented computing.



Beyond the HPC systems, the rapid development of cloud environment keeps accepting increasing amount of tasks. Consequently, such quick ramp up leads to a critical energy consumption concern~\cite{dce1,dce2,islamhpca2016} in modern data centers. Cloud service providers have widely adopted the throughput-oriented computing as one of the most effective manners to safe energy consumption, as it can consolidate/stack multiple tasks on the shared computing resources and leave unused machines power off.
Prior studies~\cite{isca2013cook,cooper,bubbleup,bubbleflex} have shown that such task consolidation can significantly improve hardware utilization and result in high energy efficiency.

Clearly, the goals of task consolidation are to: (1) increasing the system throughput and (2) improving hardware utilization and energy efficiency. However, efficiently application co-running in modern data centers is challenging. From the hardware perspective, there are many shared resources, such as last-level cache (LLC) and memory bandwidth, which can be affected by these co-running applications, leading to significant contention~\cite{dingchencgo2013,isca2013cook,lingjiaisca11}. From the software perspective, there are applications in various domains, such as data mining, machine learning, standard benchmarks, HPC, and real-world parallel applications, among others. Applications from different domains can have different requirements (e.g. Quality of Service (QoS)) and different characteristics that can lead to unpredictable co-running behaviors~\cite{bubbleup,bubbleflex,chenatc2017}. Given a large variety of individual application's characteristics, a high throughput-oriented system faces the challenges of understanding and improving task consolidation.


To overcome such challenges, prior work proposed solutions that fall into two categories: task scheduling and code compilation. Task scheduling techniques~\cite{raojia2016,bubbleflex,Wang2013,chenatc2017,wangipdps2015} preempt the applications with non-critical performance requirement and prioritize the QoS-sensitive ones in the contentious period. Code optimization techniques~\cite{lingjiacgo2012,reqos,dingchencgo2013} throttle an application's memory access intensity (e.g., reduce memory instruction issue rate or cache requirement), which results in less interference in shared resources. Both methods heavily rely on the characterization of interference. However, existing approaches~\cite{marshipeac11,isca2013cook,lingjiaisca11} provided interference characterizations of either traditional benchmarks (e.g. SPEC CPU2006~\cite{spec2006}) or a single domain (e.g. query-based web service), which provides limited insights into co-running interference of modern applications across various domains. 
In this paper, we conduct an empirical study to address this issue in prior work.

We summarize our contribution as follows:



\begin{itemize}
\item To conduct a comprehensive evaluation of co-running environments, we select 25 emerging applications/benchmarks, including graph analytics, deep learning applications, standard CPU/memory benchmarks, parallel benchmarks, and HPC applications. Most of them are up-to-date ones, such as GeminiGraph~\cite{geminigraph}, SPEC CPU2017~\cite{spec2017}, and Microsoft CNTK~\cite{cntk}.
\item We leverage the selected applications to form 625 consolidated task  pairs. The evaluation of each pair is performed three times to obtain the accurate runtime and interference measurement. 
\item To understand each application's characteristics, we analyze its scalability, prefetching sensitivity, and memory bandwidth consumption in the sole-run. We then measure each co-running pair's runtime and bandwidth usage to truly quantify the interference effect caused by task consolidation. For heavily affected applications, we leverage mini-benchmarks and real applications to deep dive into the provenance of performance degradation from both hardware and software aspects.
\end{itemize}

We have observed many insightful findings via our experiments. For instance, we find that graph analytic applications are vulnerable to cache and memory contention; they does not degrade their co-runners but can be harmed seriously by memory intensive applications. Moreover, we identify contentious code regions under severe contention, which can benefit software design for minimizing co-running interference. We believe our thorough analysis can benefit future researches from multiple domains.

The rest of the paper is organized as follows. Section~\ref{sec:related} reviews the related prior work. Section~\ref{sec:exp} illustrates our platform configuration, describes selected workloads, and introduces the tools used in experiments. Application's scalability, prefetching sensitivity, and bottlenecks are discussed in Section~\ref{sec:nointerference}. Then, the results of comprehensive consolidation experiments are presented in Section~\ref{sec:interference}. Based on these result, we deep-dive to analyze the provenance of interference in Section~\ref{sec:analysis}. Finally, we conclude this paper in Section~\ref{sec:conclusion}.

%% file: related.tex
Before we demonstrate our evaluation results, we review the state-of-the-art works that characterize interference in task consolidation environments, and minimize interference via scheduling or compilation techniques.
\subsection{Interference Characterization}
To solve the interference issues (e.g., resource contention), a comprehensive analysis is always necessary. Mars et al.~\cite{marshipeac11} leveraged  a synthesis engine to characterize interference sensitivity of SPEC CPU2006 benchmarks and identify contentious application code. Moreover, Tang et al.~\cite{lingjiaisca11} studied the impacts of sharing memory resources on five Google applications (protocol buffer, web search, etc.) and leverage the analysis results to guide thread-to-core mappings for better performance. Cook et al.~\cite{isca2013cook} evaluate the potential hardware cache partitioning schemes for multi-tenant executions formed by PARSEC~\cite{parsec} and SPEC CPU2006~\cite{spec2006} applications. Bubble-up~\cite{bubbleup} presents a characterization methodology that can predict latency-sensitive tasks' (e.g. web search) performance degradation caused by shared memory subsystems. To better understand the resource sharing in servers, Dong et al.~\cite{dong2016venice} designed Venice, a hardware prototype, to transparently monitor the consolidation interference at runtime. Other than the interference analysis work, Delimitrou et al.~\cite{ibench} released the iBench benchmark suite, which consists of manual-crafted benchmarks that can stress different hardware components, such as CPU, cache hierarchy, memory, etc.

\subsection{Scheduling}
Leveraging the comprehensive characterization of application's performance in multi-tenant environments, many researchers have proposed to minimize the interference effects via scheduling techniques. 
Zhao et al.~\cite{raojia2016} proposed delayed preemption and differential scheduling to interleave foreground/background tasks and avoid the co-location case of interfering workloads. Bubble-flux~\cite{bubbleflex} monitored the QoS bounded applications and schedule the background tasks to guarantee the QoS quality of foreground tasks. By training a simple program classifier, Wang et al.~\cite{Wang2013} created an interference-aware schedule to avoid contentious application pairings. To achieve fair sharing in cloud computing, Chen et al.~\cite{chenatc2017} designed two preemption policies (\textit{preemptive} and \textit{graceful}) for the container-based frameworks. Cooper~\cite{cooper} leveraged the stable matching algorithm to guide task consolidation. 
Wang et al.~\cite{wangipdps2015} presented CC, a scheduling algorithm that leveraged the benefits of contention while supporting limited reservations to reduce interference with fairness guaranteed. 

\subsection{Compilation}
Besides the scheduling techniques, Tang et al.~\cite{lingjiacgo2012} proposed the first compilation approach to statically pinpoint and throttle the memory access rate of an application's contentious region, which resulted in less interference in the co-running environment. Similarly, ReQos et al.~\cite{reqos} leveraged a profile-guided compilation technique to insert marks in contentious code regions. Marked code would reactively reduce pressure on memory subsystems when contention was detected. To solve the inclusion victim problem in co-running environments, Bao et al.~\cite{dingchencgo2013} used defensive tiling to estimate cache sharing effect and then select optimal tile size that can minimize interference in cache.

%% file: experiment.tex
In this section, we provide an overview of the experimental platform, describe the applications in detail, and present two toolsets used for bandwidth measurement and bottleneck detection.

\subsection{Platform Configuration}
Our experiments are performed on a Supermicro 8047R-TRF+ server with 8-core Intel Xeon E5-4650 processor clocked at 2.7GHz. Each core has a private 32K L1 instruction Cache, 32K L1 data Cache and 256K L2 Cache. All cores share a 20MB L3 Cache and 64 GB DRAM. The operation system on this machine is CentOS 7.

\begin{figure}[t]
  \centering
  \includegraphics[width=3in]	       {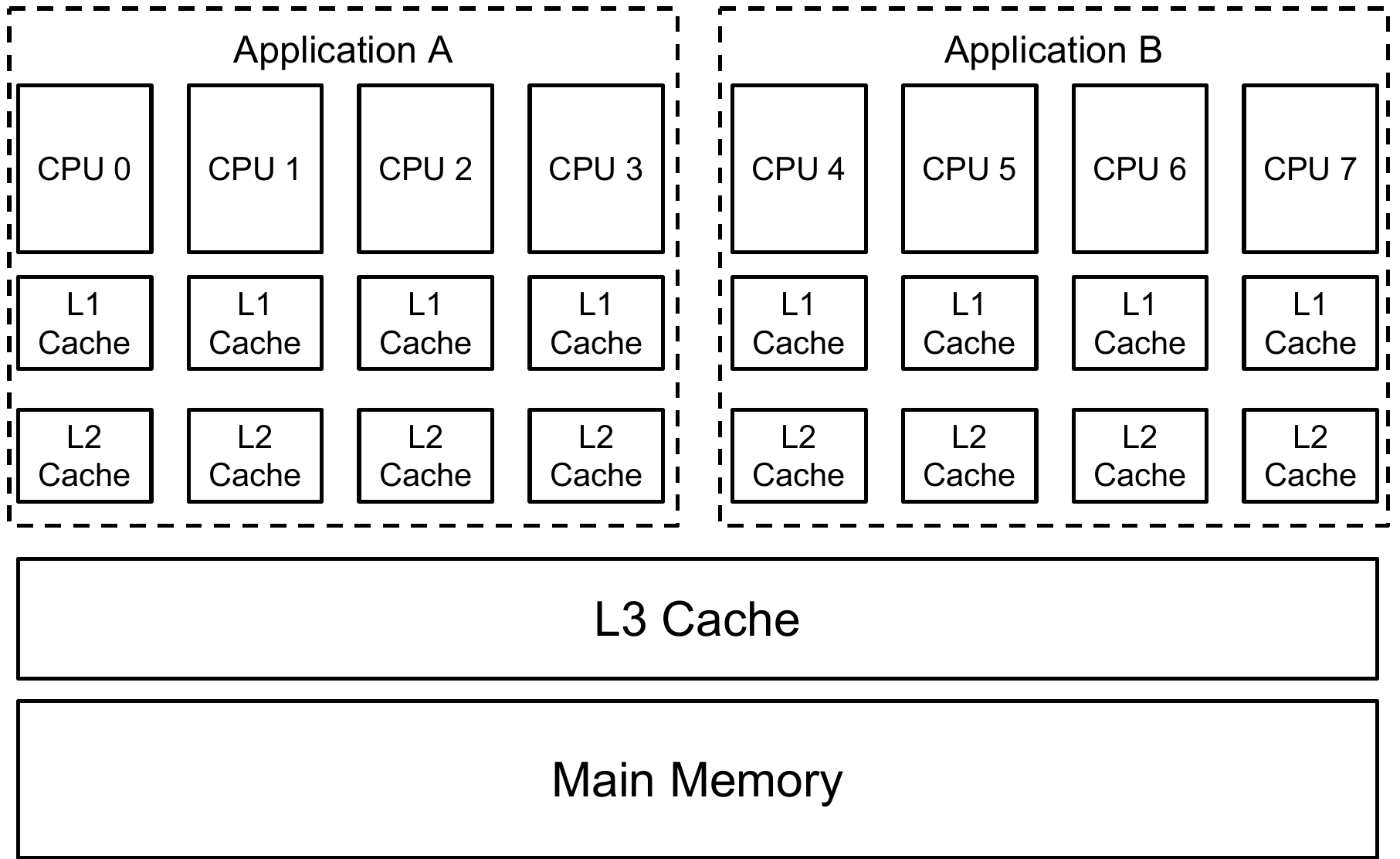}
  \caption{Experiment Setup}
  \label{fig:experiment_setup}
  \vspace{-5mm}
\end{figure}

Similar to the server configuration in warehouse-scale computing~\cite{tacc1,tacc2,amazonht}, we disable the Hyper-Threading (HT) technology of the machine, as HT can reduce certain application's performance. 
Furthermore, to minimize the effects (e.g. context switch) on applications performance, we binded each application to specific cores exclusively. Figure \ref{fig:experiment_setup} shows the binding configurations for the co-running experiments. All applications are configured with 4 threads. Each application takes 4 physical cores exclusively. With such configuration, the only shared resources are the LLC and the memory subsystem.

\subsection{Workload Description}

We form our workload set using a wide range of applications from 5 application domains: Graph Processing, Deep Learning, SPEC CPU2017, PARSEC and HPC. Table \ref{tab:benchmark_suite} summarizes the chosen applications. Details of each application suite are summarized as follows:
\paragraph{\textbf{PowerGraph}} PowerGraph~\cite{powergraph} is one of the earliest graph systems that is designed for data mining and analytics. It is implemented in the classic bulk synchronization manner. Techniques used in it have been widely adopted by many other graph frameworks~\cite{powerswitch,powerlyra,cube,shuangsc,shuangicpp}. 

\paragraph{\textbf{GeminiGraph}} GeminiGraph~\cite{geminigraph} is one of the state-of-the-art graph processing frameworks with near-linear scalability. This framework emphasizes on optimizing computation performance via thread-level partitioning and balancing. We choose 5 representative data analytics applications implemented atop it, which include Pagerank (PR), single source shortest path (SSSP), breadth first search (BFS), betweenness centrality (BC), and connected component (CC). We use the \textit{friendster} graph~\cite{snapnets} with 65.6 million vertices and 1.8 billion edges in both GeminiGraph and PowerGraph.

\paragraph{\textbf{CNTK}} CNTK (short for The Microsoft Cognitive Toolkit)~\cite{cntk} is an open-source and commercial grade toolkit of deep learning applications. We deploy 3 applications from this toolkit. ConvNet is an image recognition workload with CIFAR and MNIST datasets. LSTM is a Long Short Term Memory network using CMU-AN4 data input. ATIS a natural language processing workload, which leverages the data from the air travel information system. For these deep learning applications, only the training phase is used for performance measurement.

\paragraph{\textbf{PARSEC}} The PARSEC benchmark suite~\cite{parsec} is used to represent parallel real-world applications. We use the largest input sets that are designed for native execution. Four benchmarks are chosen from this suite, which are blackscholes, freqmine, swaptions, and streamcluster. 

\paragraph{\textbf{SPEC CPU2017}} The SPEC CPU2017 benchmark suite is recently released to represent CPU/memory-intensive applications. It is designed to mimic the real-world applications. Using the subseting methodology suggested by Phansalkar et al.~\cite{Phansalkar2007}, we obtain the performance data (published on SPEC website~\cite{spec2017}) submitted by industries and choose 6 benchmarks to represent SPEC CPU2017 suite. CactuBSSN, nab, and fotonik3d are floating-point benchmarks. Xalancbmk, mcf, and deepsjeng are integer benchmarks. To achieve the effect of parallel execution (SPEC rate), multiple copies of each workload are executed simultaneously. 

\paragraph{\textbf{HPC}} Three benchmarks published by Lawrence Livermore National Laboratory are chosen: AMG2006, IRSmk and lulesh. AMG2006 is a parallel algebraic multi-grid solver for linear systems. IRSmk is a parallel nested do-loops that perform a matrix multiply. Lulesh solves the Sedov blast wave problem for one material in 3D.

Other than these real-world applications, we leverage two mini-benchmarks to study specific memory system features such as bandwidth. The two mini-benchmarks are Bandit\cite{haoxu2017} and Stream \cite{stream1995}, both of which are memory stressing benchmarks. The Bandit program continuously issues memory requests, where all requests result in cache misses. Every memory access in Bandit conflicts with its previous one in caches, so such access goes to the main memory. Stream program streams a large amount of data from memory. It has a very regular memory access pattern. Hence, hardware prefetchers help load consecutive cache lines from memory, which results in extremely high memory bandwidth consumption.

\begin{table}[t]
\centering
\caption{Applications chosen for each Application Suite}
\label{tab:benchmark_suite}
\begin{tabular}{|c|c|}
\hline
Benchmark Suite & Benchmarks \\ \hline
GeminiGraph & PR, BFS, BC, SSSP, CC \\ \hline
PowerGraph & PR, SSSP, CC \\ \hline
CNTK & ConvNet-CIFAR/MNIST, LSTM-AN4, ATIS\\ \hline
PARSEC & freqmine, streamcluster, swaptions, blackscholes \\ \hline
HPC & lulesh, amg2006, IRSmk \\ \hline
SPEC CPU2017 &  mcf, fotonik3d, deepsjeng \\ 
& nab, xalancbmk, cactuBSSN \\ 
\hline
mini-benchmarks & Bandit, McCalpin Stream \\ \hline
\end{tabular}
\end{table}

\subsection{Performance Measurement}
To measure the bandwidth consumption at runtime, we adopted Intel Processor Counter Monitor (PCM) 2.8~\cite{pcm} to read MSR registers. PCM provides a set of tools based on its API to monitor performance and energy metrics of Intel processors. We used pcm-memory to measure bandwidth at 10 seconds granularity with extremely low overhead.
In addition, we used Intel VTune Amplifier 2017~\cite{vtune} to find performance bottleneck of problematic co-running pairs. VTune can provide hardware event-based sampling analysis result, with a very low overhead. Beside, it also can attribute event sampling results to specific hot spots.

%% file: run_no_interference.tex
In this section, we do experiments to explore each application's thread scalability, bandwidth consumption, prefetcher sensitivity, without co-running. 

\begin{figure*}
    \centering
    \begin{subfigure}[b]{0.3\textwidth}
        \includegraphics[width=\textwidth]{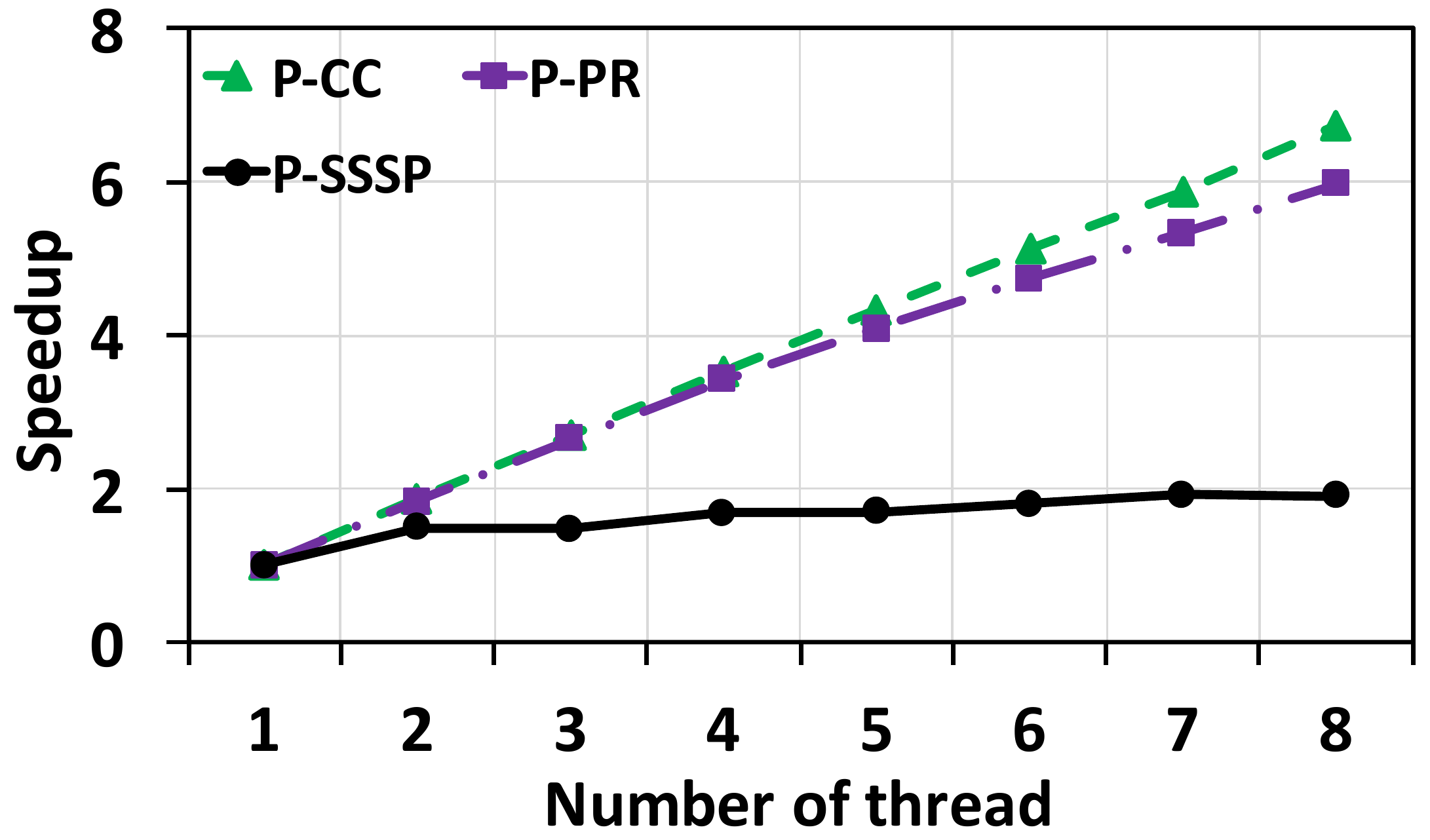}
        \caption{Powergraph}
        \label{fig:scale_powergraph}
    \end{subfigure}  
    \begin{subfigure}[b]{0.3\textwidth}
        \includegraphics[width=\textwidth]{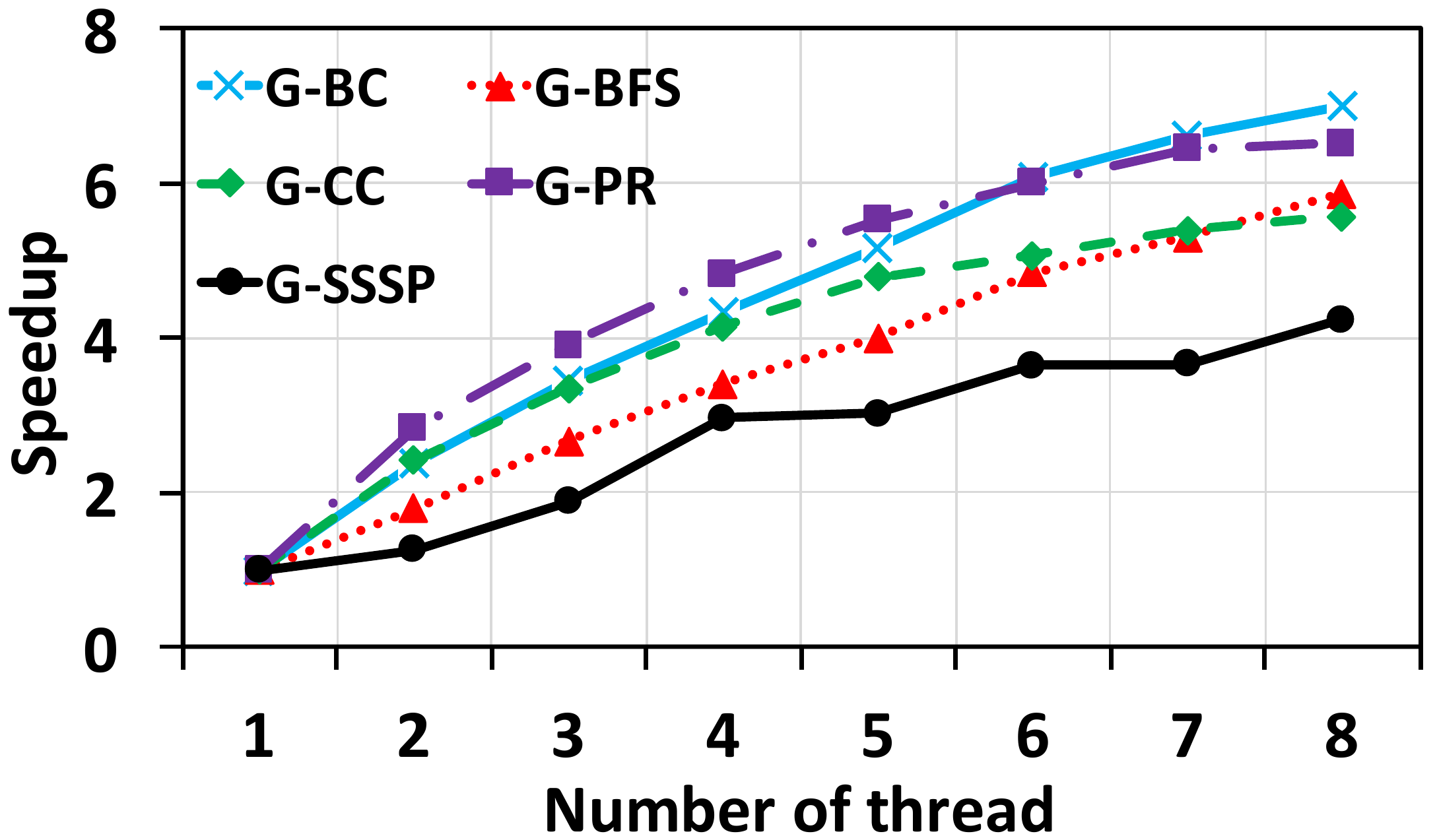}
        \caption{Gemini}
        \label{fig:scale_gemini}
    \end{subfigure}
    \begin{subfigure}[b]{0.3\textwidth}
        \includegraphics[width=\textwidth]{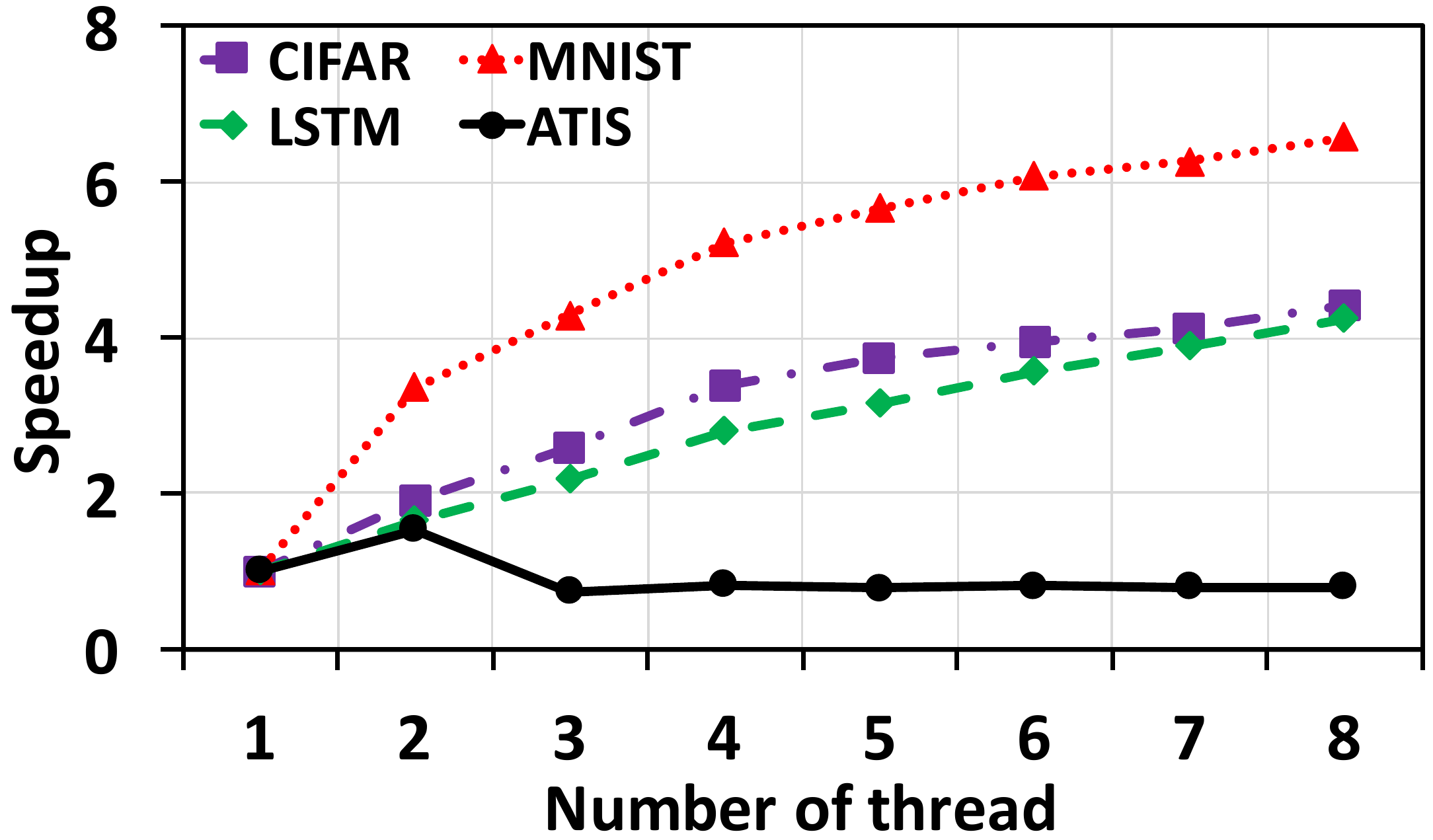}
        \caption{CNTK}
        \label{fig:scale_cntk}
    \end{subfigure}
    
    \begin{subfigure}[b]{0.3\textwidth}
        \includegraphics[width=\textwidth]{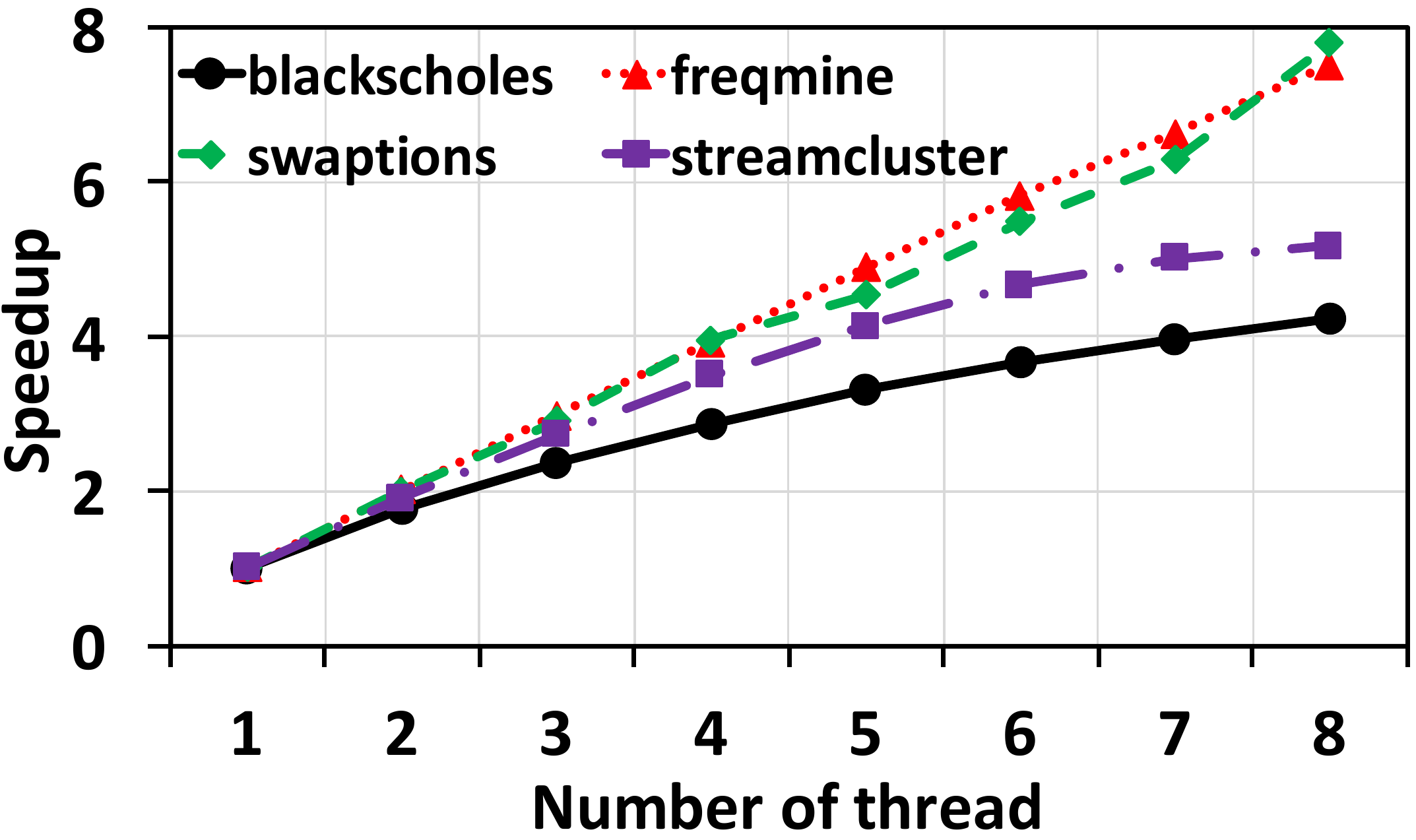}
        \caption{PARSEC}
        \label{fig:scale_parsec}
    \end{subfigure}
    \begin{subfigure}[b]{0.3\textwidth}
        \includegraphics[width=\textwidth]{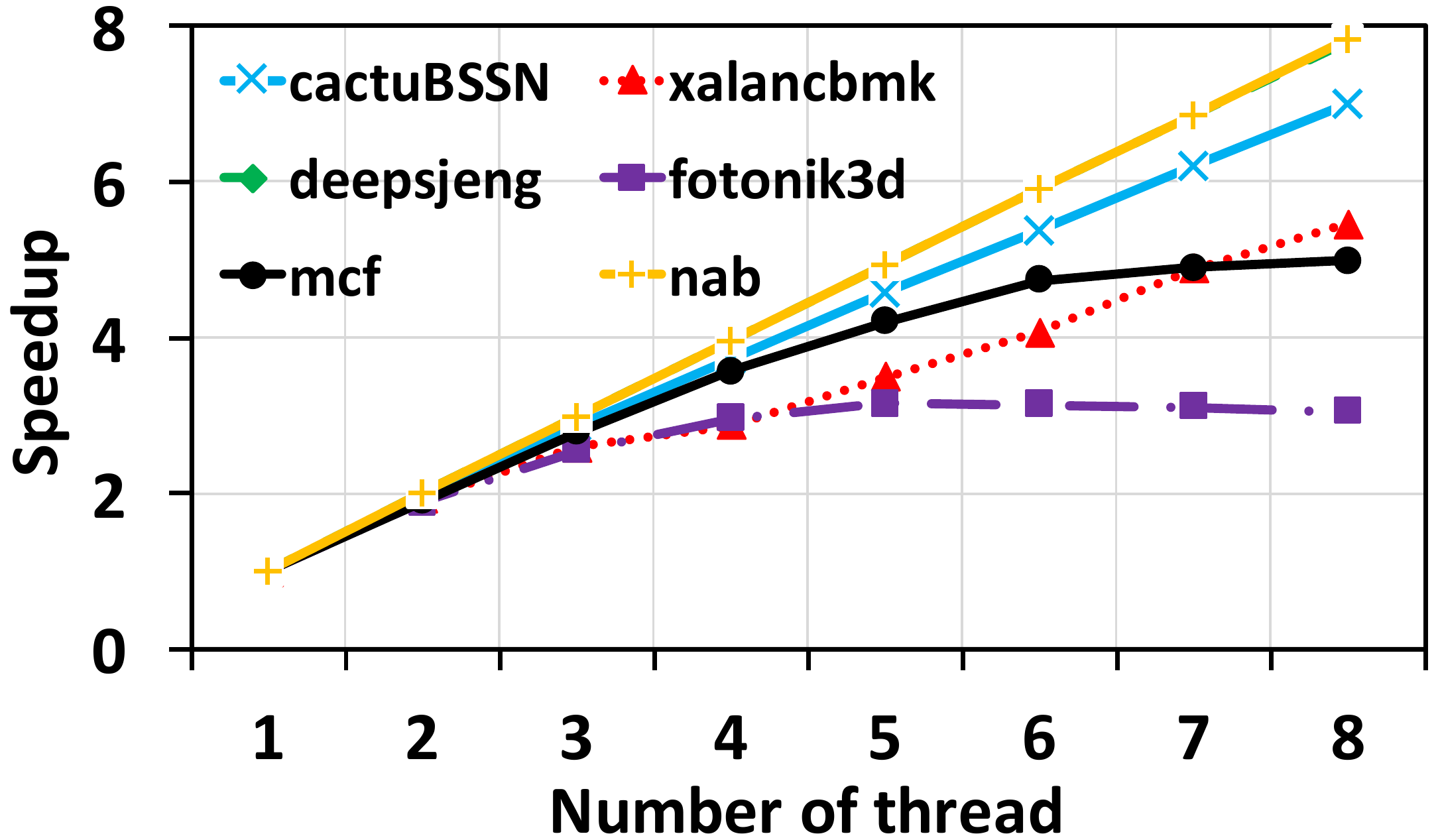}
        \caption{SPEC CPU2017}
        \label{fig:scale_spec}
    \end{subfigure}
     \begin{subfigure}[b]{0.3\textwidth}
        \includegraphics[width=\textwidth]{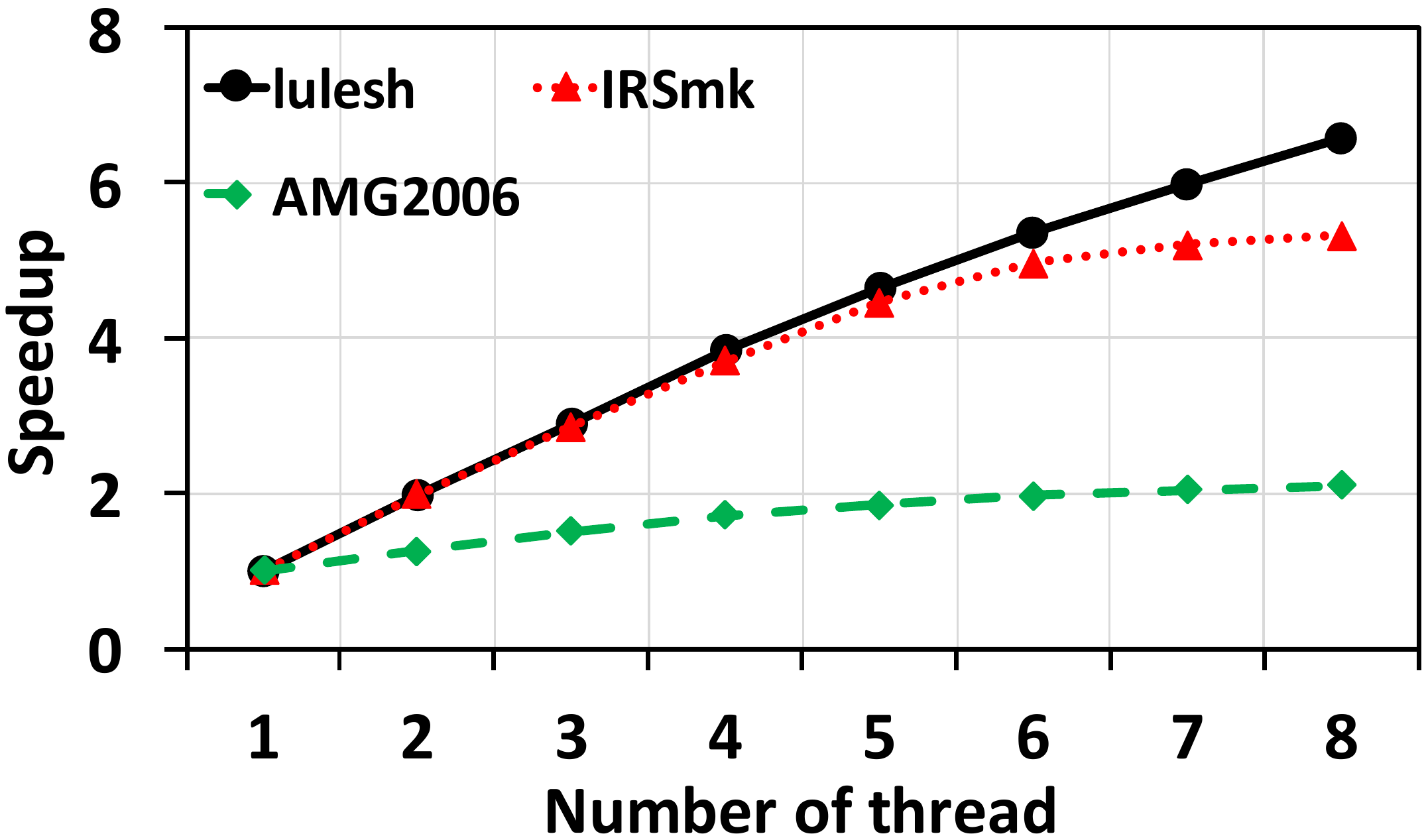}
        \caption{HPC}
        \label{fig:scale_hpc}
    \end{subfigure}
    \caption{Normalized speed up for different number of threads of each application.}
    \vspace{-5mm}
\end{figure*}

\subsection{Thread Scalability}
\label{sec:threadscale}
To evaluate each application's thread scalability, we run each application on this machine exclusively with fixed input size. Each application's thread number increases linearly from 1 to 8. 
The application's execution time is measured to calculate the speedup among different thread counts. 
For application with a non-negligible data preprocessing phase, we only measure the runtime of the execution phase. 
For instance, using 4 threads to preprocess the \textit{friendster} graph in GeminiGraph can take approximately 420s, while executing the SSSP application atop it only takes 48.6s. Similar phenomenon can be found in PowerGraph as well.  
Such preprocessing time is not included in the execution time, as this is a one-time cost and it can be amortized with multiple runs.

\begin{figure}[t]
  \centering
  \includegraphics[width=3.5in,trim = 0mm 0mm 0mm 0mm, clip=true]	   
   {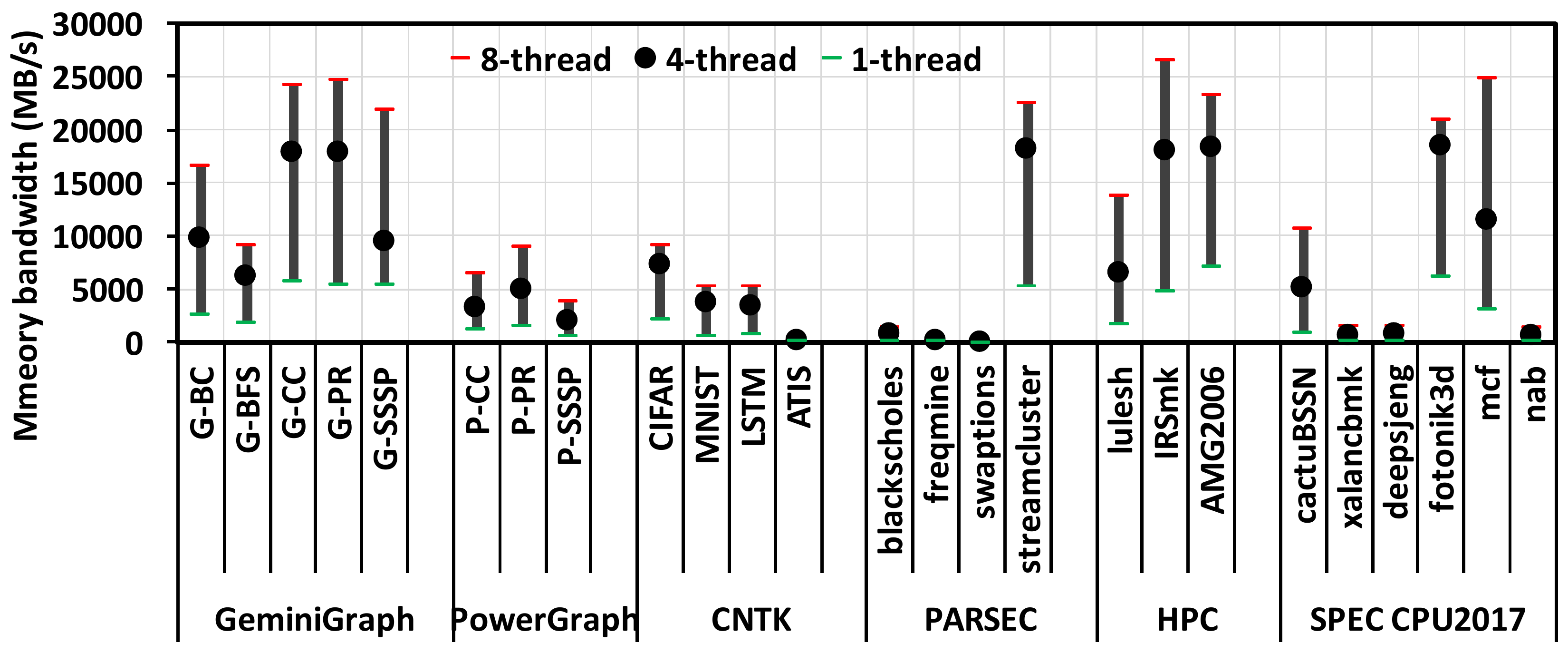}
  \caption{Memory bandwidth of each application.}
  \label{fig:bandwidth_eachapp}
  \vspace{-5mm}
\end{figure}

As Figure \ref{fig:scale_powergraph} shows, the PowerGraph applications' scalability is good except P-SSSP: P-CC and P-PR scale linearly up to $6.7\times$, P-SSSP shows speedup less than $2\times$. The low scalability of P-SSSP is caused by the unrealistic assumption that all graph edges have identical weight. Its low scalability is observed in~\cite{shuangicpp} as well. GeminiGraph applications scale well(shown in Figure \ref{fig:scale_gemini}): all the applications can get more than $4\times$ scaling factors, and the maximum is up to $7\times$. G-PR, G-CC and G-BC grow faster when the number of threads is less than 4, and then their scale factor growth slow down. As shown in Figure~\ref{fig:bandwidth_eachapp}, these three applications' bandwidth are relatively higher and nearly saturated at 4-thread configuration. 
Compared to other graph applications, the speedup of G-SSSP on this scale-up test is less sharp. 
This phenomenon is due to the nature of G-SSSP's irregular memory access pattern~\cite{burtscher2012,neil2014} rather than the bandwidth bottleneck. Moreover, CNTK applications scalability varies a lot(Figure \ref{fig:scale_cntk}): CIFAR, MNIST and LSTM have good scalability up to $6.3\times$, while ATIS has no scalability. We use the Intel Vtune to profile ATIS and find that ATIS spends 80\% of total CPU cycles on \texttt{kmp\_hyper\_barrier\_release} function (synchronization bound), when its thread number is above 2. However, this function only costs 28\% execution cycles in 2-thread case. 

PARSEC applications have good scalability with more than $4\times$ speedup, especially  blackscholes and freqmine's speedup are nearly $8\times$. Furthermore, SPEC CPU2017 applications' scalability results are plotted in Figure \ref{fig:scale_spec}. CactuBSSN, nab and deepsjeng have perfect linear scalability. However, fotonik3d scales poorly after $4$ threads. Since fotonik3d is a memory intensive application, 4 fotonik3d instances cause memory bandwidth to reach its limit in practice. Therefore, the scalability of fotonik3d is relatively low. Finally, Figure \ref{fig:scale_hpc} shows HPC application's scalability: lulesh scales well, and IRSmk and AMG2006 saturates after 6 threads and 4 threads, respectively. AMG2006 has three different phases, where the first two phases leverage single thread to process input data. In its last phase, it generates intensive memory accesses and yields a medium scalability.

Based on each application's thread scalability performance, we characterize all applications into three categories: low scalability, saturate scalability, and high scalability. Table \ref{tab:thread_scale_cat} presents the thread scalability category for each application. 
Most graph processing applications (data analytics) have high scalability. For CNTK (machine learning) and HPC workloads, the scalability highly depends on the structure of the application. For  SPEC CPU2017 and PARSEC benchmarks, they either saturated around 4-thread or yields high scalability.
The information provided by our analysis can help choose the right configuration for execution without wasting hardware resources in the cloud environment.

\begin{table}[t]
\centering
\caption{Thread scalability characterization result.}
\label{tab:thread_scale_cat}
\begin{tabular}{|c|c|c|c|}
\hline
Suite & Low & Medium & High \\ \hline
Powergraph & P-SSSP & - & P-CC, P-PR \\ \hline
Gemini & - & G-SSSP & \begin{tabular}[c]{@{}c@{}}G-CC,G-BC,\\ G-PR,G-BFS\end{tabular} \\ \hline
CNTK & ATIS & CIFAR, LSTM & MNIST \\ \hline
PARSEC & - & \begin{tabular}[c]{@{}c@{}}streamcluster,\\ blackscholes\end{tabular} & swaptions,freqmine \\ \hline
SPEC CPU2017 & - & \begin{tabular}[c]{@{}c@{}}fotonik3d, \\ deepsjeng,cpuxalan\end{tabular} & \begin{tabular}[c]{@{}c@{}}mcf, nab,\\ cactuBSSN\end{tabular} \\ \hline
HPC & AMG2006 & IRSmk & lulesh \\ \hline
\end{tabular}
\vspace{-5mm}
\end{table}

\subsection{Bandwidth Analysis}

Figure \ref{fig:bandwidth_eachapp} shows each application's bandwidth consumption range with different number of threads, measured by Intel PCM 2.8. Due to the page limit, we only show bandwidth consumption for three configurations: 1-thread, 4-thread, 8-thread, where 1-thread yields the minimal bandwidth usage and 8-thread produces the maximal one.

Similar to the runtime measurement, for graph applications, the bandwidth information is measured in every execution phase. GeminiGraph applications consume higher bandwidth than PowerGraph applications, as GeminiGraph leverages thread-level work stealing technique and achieves better data locality due to the chunking partitioning scheme. Streamcluster in PARSEC, IRSmk and AMG2006 in HPC, fotonik3d and mcf in SPEC CPU2017 consume a larger amount of bandwidth compared to other applications in the same domain. Since we measure bandwidth consumption of each application with different thread configurations, we can verify the correlation between the bandwidth consumption and thread counts. Other than the graph applications discussed in Section~\ref{sec:threadscale}, we find that streamcluster, IRSmk, fotonik3d and CIFAR's bandwidth consumption growth becomes much slower as thread counts scales up from 4 to 8. The scalability of these benchmarks also saturates after 4-thread. We conclude that these benchmarks don't benefit from the increase in thread resources after 4 threads, as they are bounded by other resources, like the memory bandwidth. Workloads, like ATIS, blackscholes, freqmine, swaptions, xalancbmk, deepsjeng and nab, have extremely low bandwidth consumption. 

\subsection{Prefetcher Sensitivity}

Next we study each application's sensitivity to hardware prefetchers, as prefetcher is one of the most effective techniques for data intensive applications. Workloads with regular memory access pattern can benefit a lot from hardware prefetchers. Such applications usually take a lot of bandwidth resource and potentially degrade other co-running application's performance. This sensitivity study can help us identify such applications.

For Sandy Bridge machine,  there are four different hardware prefetchers: 
\begin{itemize}
\item L2 hardware prefetcher: fetches additional cache lines into L2 cache.
\item L2 adjacent cache line prefetcher: fetches successive cache lines.
\item L1-data cache prefetcher: fetches the next cache line into L1-D cache.
\item L1-data cache IP prefetcher: fetches additional cache lines determined by instruction point of previous load history.
\end{itemize}
For each core, there is a Model Specific Register (MSR) to control hardware prefetchers. Each prefetcher can be activated or deactivated by setting the corresponding MSR bit. Such MSR bits can be found on Ivy Bridge, Haswell, Broadwell processors as well~\cite{intelmanual}.

\begin{figure}[t]
  \centering
  \includegraphics[width=3.5in,trim = 0mm 0mm 0mm 0mm, clip=true]	     
   {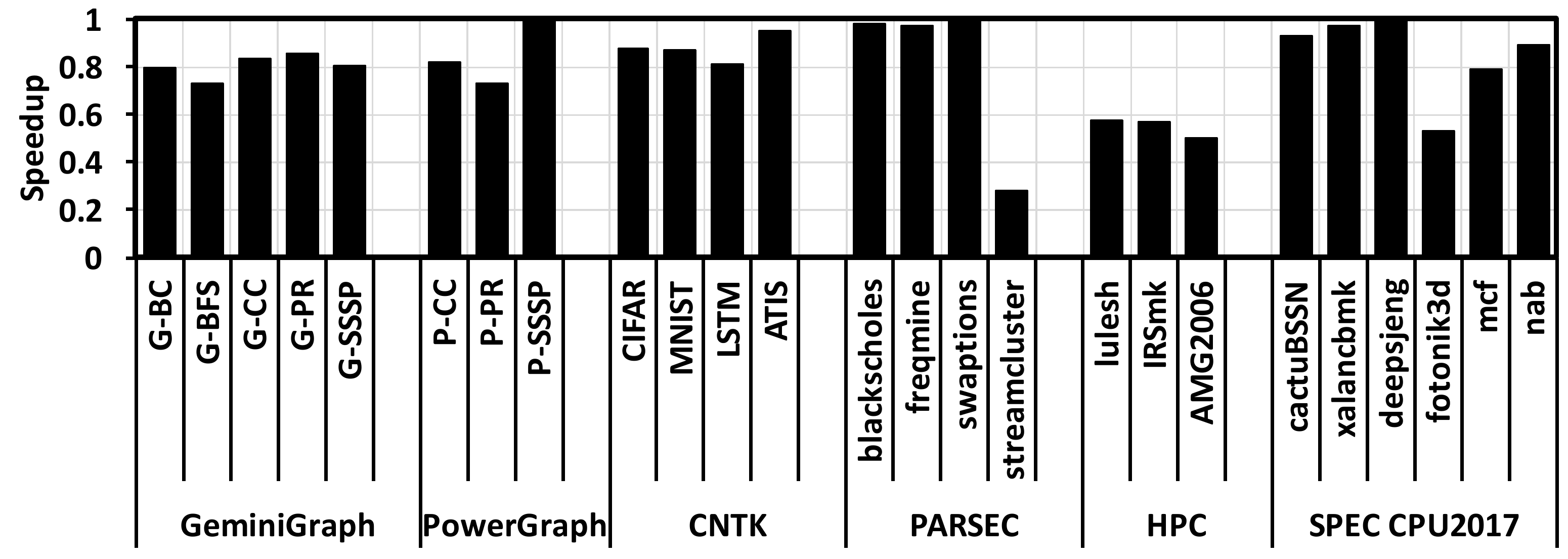}
  \caption{Prefetch sensitivity: Slowdown if prefetcher is turned off}
  \label{fig:prefetch_sensitivity}
  \vspace{-5mm}
\end{figure}
We fix each application's thread-count and use 4-thread with prefetchers enabled as our baseline. 
Each application's execution time with all prefetchers activated is normalized to the execution time with prefetchers deactivated. Results are plotted in Figure 4.
In GeminiGraph and PowerGraph, all graph applications do not benefit from L1/L2 cache prefetchers, as their memory access patterns are not regular enough. Similar to graph applications, machine learning applications in CNTK are not sensitive to prefethers. However, the root cause can be different, as the memory bandwidth consumption of graph applications is $2.45\times$ higher than CNTK applications'. Therefore, applications like ATIS is very robust to the availability of prefetchers. Such applications are unlikely to harm other consolidating application.

\begin{figure*}[t]
  \centering
\includegraphics[height=0.425\textheight, width=0.85\textwidth, clip=true]
   {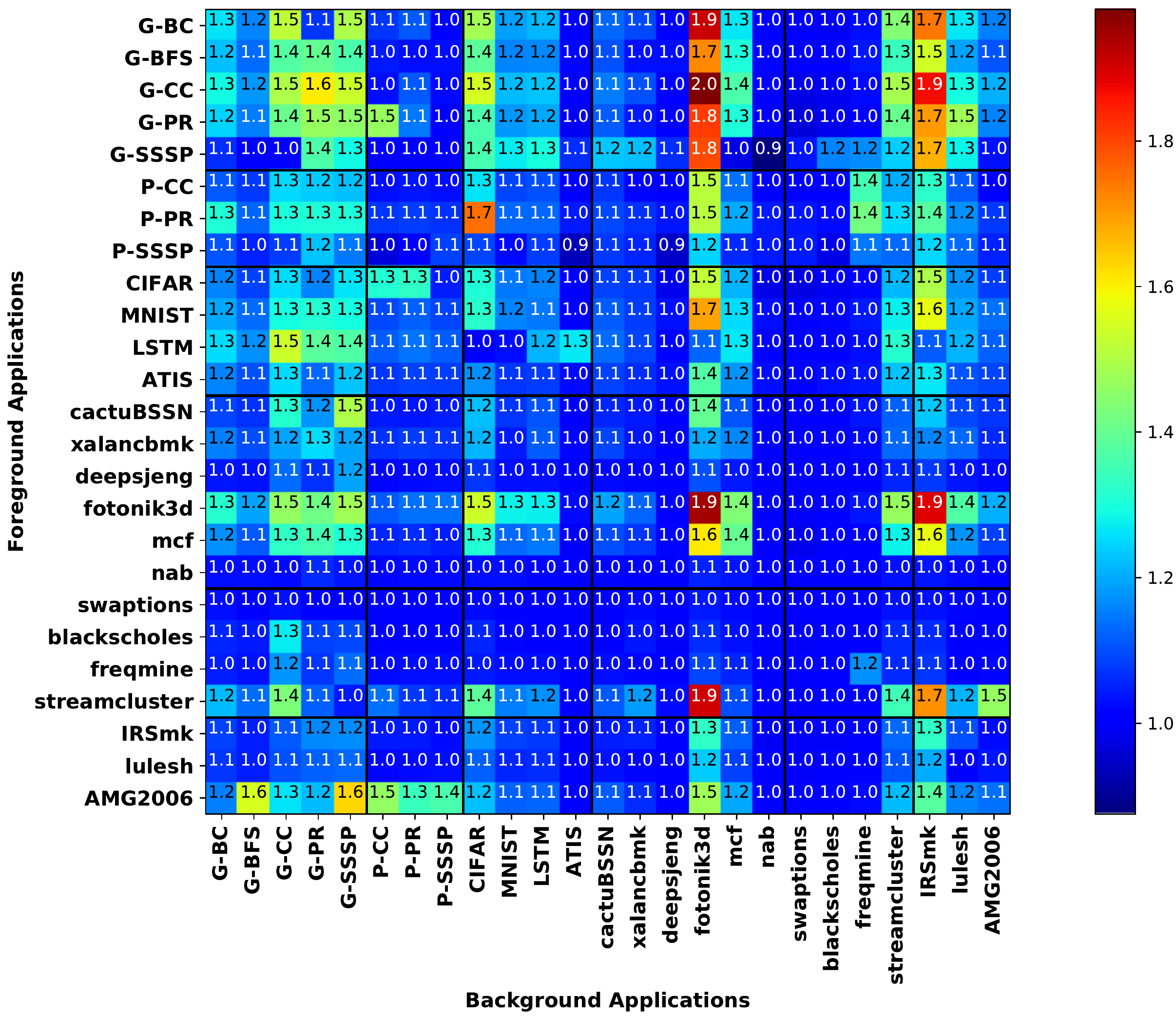}
  \caption{Normalized execution time of co-running two applications on the shared machine (foreground application on y-axis and background application on x-axis).}
  \label{fig:corun_heatmap}
  \vspace{-4mm}
\end{figure*}

The streamcluster in PARSEC,  HPC applications, and fotonik3d in SPEC CPU2017 are very sensitive to prefetchers, as their performance is slowed down by 1.18$\times$. All these benchmarks have a regular memory access pattern, and consume significant amount of bandwidth. These applications can potentially cause LLC and memory contention problems in the co-running case.

%% file: run_interference.tex
After analyzing each application solely, this section presents our analysis of each application in the co-running environment. For each co-running pair, we classify the two applications as follows.
\begin{itemize}
\item {\bf Background application}: the application executed in background infinitely when foreground benchmark is running. It will be stopped at once, when foreground benchmark stops.
\item {\bf Foreground application}: the application executed in foreground, we measure its execution time. 
\end{itemize}

To form a fair sharing setup, each applications' thread number is set to 4. Each thread takes one physical core exclusively.  
Figure \ref{fig:corun_heatmap} shows a heat map of the normalized execution time of each foreground application to the execution time of the application running without the interfering neighbor. The application shown on y-axis is the foreground task and one shown on x-axis is the corresponding background one.

Based on the change of application's runtime, we define three kinds of relationship for application $A$ and $B$ in a co-running pair:
\begin{itemize}
\item \textbf{Harmony}: Both $A$ and $B$ in the co-running pair performs well in multi-tenant execution. No matter $A$ or $B$ is treated as foreground application, its relative execution time is $<1.5\times$.
\item \textbf{Victim-Offender}: The runtime of application $A$ in co-running pair is increased by $\ge 1.5\times$, while $B$ performs relatively well (runtime increment $<1.5\times$). 
\item \textbf{Both-Victim}: The performance of both application $A$ and $B$ are severely impacted. 
Compared to the sole-run, both $A$'s and $B$'s runtime increase $\ge 1.5\times$.
\end{itemize}

In the cloud computing or warehouse-scale computing, the \textbf{Harmony} relationship is the most preferable one, as it can significantly increase the hardware utilization rate, which results in higher overall throughout and energy efficiency. \textbf{Victim-Offender} is also acceptable, as long as the foreground task maintains its performance at a pre-defined level (not a victim). \textbf{Both-Victim} should be definitely avoided. 

\subsection{Performance Degradation of Co-running Pairs}
As shown in Figure \ref{fig:corun_heatmap}, we can find most of the co-running pairs are in a \textbf{Harmony} relationship. Especially, when swaptions, nab, deepsjeng and blackscholes are treated as background applications, any foreground applications' execution time increases less than $10\%$. Besides, no matter which background application is chosen to co-run with these benchmarks, their execution time is affected very little (less than $10\%$). These benchmarks are more ``friendly" for the shared-resource environment. From bandwidth perspective (shown in Figure \ref{fig:bandwidth_eachapp}), the bandwidth consumption of these workloads is much less than the overall bandwidth resource the whole system can supply. 

We also find several \textbf{Victim-Offender} co-running pairs: G-CC with CIFAR, G-CC with fotonik3d, P-PR with fotonik3d, IRSmk with fotonik3d, etc. 
For example, when G-CC is co-locating with CIFAR, the execution time of G-CC is increased by nearly $54.7\%$ as compared to its runtime in no-interference case. 
However, CIFAR's slowdown is only $25\%$. Therefore, we can clearly conclude that G-CC is the victim and CIFAR is the offender in this pair. When G-CC is running with fotonik3d, G-CC's performance is even worse, with the execution time goes up to $198\%$, while fotonik3d's execution time only grows $46\%$. That means fotonik3d incurs more interference/contention to G-CC than that to CIFAR. For all of these \textbf{Victim-Offender} co-running pairs shown in Figure \ref{fig:corun_heatmap}, we can conclude that graph analytical applications are more likely to be \textbf{Victim}. CIFAR, fotonik3d, and IRSmk are offenders frequently. We can find some hints from individual study of each benchmark from Section~\ref{sec:nointerference}. Fotonik3d and IRSmk have high memory bandwidth consumption and are very prefetcher sensitive. Therefore, we can conclude that application with a regular access pattern and high bandwidth cost is more likely to be a \textbf{Offender} in co-running environments. Note that AMG2006 is an exception here. As aforementioned, the third phase of AMG2006 consumes a large amount of bandwidth, which only lasts for a short execution period. 

Finally, our experiments also reveal couple of \textbf{Both-Victim} co-running pairs, such as CIFAR with fotonik3d. For this co-running pair, the execution time of CIFAR and fontonik3d increase by $52\%$ and $54\%$, respectively. Both of these benchmarks affect each other. Such consolidating pairs should definitely be avoided for cloud/warehouse-scale computing. 


\subsection{Bandwidth consumption of co-running pairs}
After analyzing applications' co-running performance , we leverage the Intel PCM tool to collect the bandwidth consumption of these co-running pairs.
In this paper, we are interested in two types of co-running pairs: \textbf{Victim-Offender} (e.g. foreground task is the victim) and \textbf{Both-Victim}. Because these two types of co-running pairs can either hurt user's experience by increasing foreground task's runtime or minimize the overall system throughput. To evaluate the bandwidth consumption of these two types of co-running pairs, we choose 5 co-running pairs to analyze. We also take each application's bandwidth consumption in these 5 co-running pairs for comparison. The baseline of such comparison is measured when the application is running alone. The bandwidth information is summarized in Table \ref{tab:bandwidth_corun}. The bandwidth consumption of all co-running pairs is less than the summation of two application's bandwidth consumption. 
All application in Table~\ref{tab:bandwidth_corun} are memory intensive ones, which can generate a significant amount of traffic on the bus. Pairing foreground tasks with them can diminish their responsiveness or elongate their execution time. Pairing such applications together can stress out the memory bandwidth, which results in reduced system throughput. After showing these phenomenon, we leverage the profiling tools (Vtune) to deep dive into these applications for the root cause of slow down.

\begin{table}[]
\centering
\scriptsize
\caption{Bandwidth Consumption of Specific Co-running Pairs. }
\label{tab:bandwidth_corun}
\begin{tabular}{|c|c|c|c|}
\hline
\multirow{2}{*}{Co-running Pairs} & \multicolumn{3}{c|}{Bandwidth Consumption(GB/S)} \\ \cline{2-4} 
 & Co-running pair & Application A & Application B \\ \hline
\begin{tabular}[c]{@{}c@{}}CIFAR(A) with \\ fotonik(B)\end{tabular} & 18.0 & 7.3 & 18.4 \\ \hline
\begin{tabular}[c]{@{}c@{}}IRSmk(A) with \\ fotonik(B)\end{tabular} & 24.5 & 18.1 & 18.4 \\ \hline
\begin{tabular}[c]{@{}c@{}}CC-G(A) with \\ fotonik(B)\end{tabular} & 18.6 & 17.8 & 18.4 \\ \hline
\begin{tabular}[c]{@{}c@{}}CC-G(A) with \\ IRSmk(B)\end{tabular} & 26.3 & 17.8 & 18.1 \\ \hline
\begin{tabular}[c]{@{}c@{}}CC-G(A) with \\ CIFAR(B)\end{tabular} & 18.6 & 17.8 & 18.0 \\ \hline
\end{tabular}
\end{table}

%% file: interference_analysis.tex
To identify the impact of co-running on shared hardware resource and contentious code region of the consolidating task, we deploy mini-benchmarks to study interference of each application, and present collected performance statistics in this section. 
Moreover, here we also discuss the insights obtained from comparing the profiling results of co-running with mini-benchmarks to the ones of consolidating with real applications.

\begin{figure}[t]
    \centering
    \begin{subfigure}[b]{0.5\textwidth}
        \includegraphics[width=3.5in]{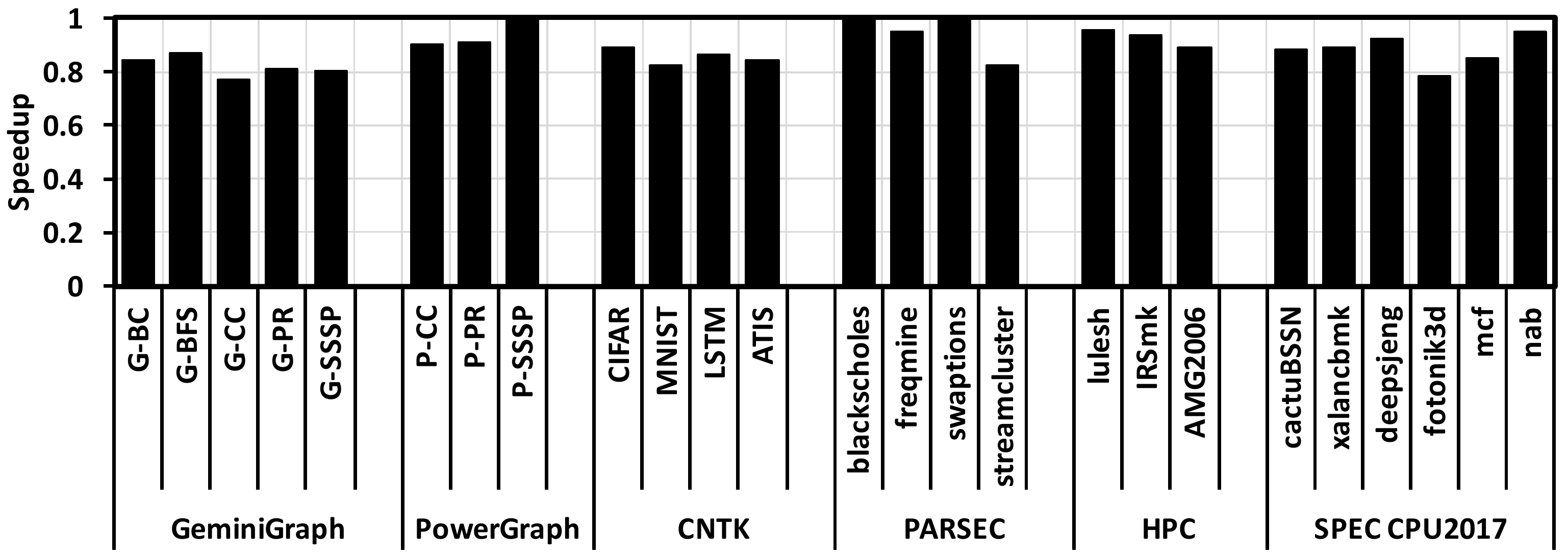}
        \caption{Co-run with Bandit}
        \label{fig:corun_stream}
    \end{subfigure}  

    \begin{subfigure}[b]{0.5\textwidth}
        \includegraphics[width=3.5in]{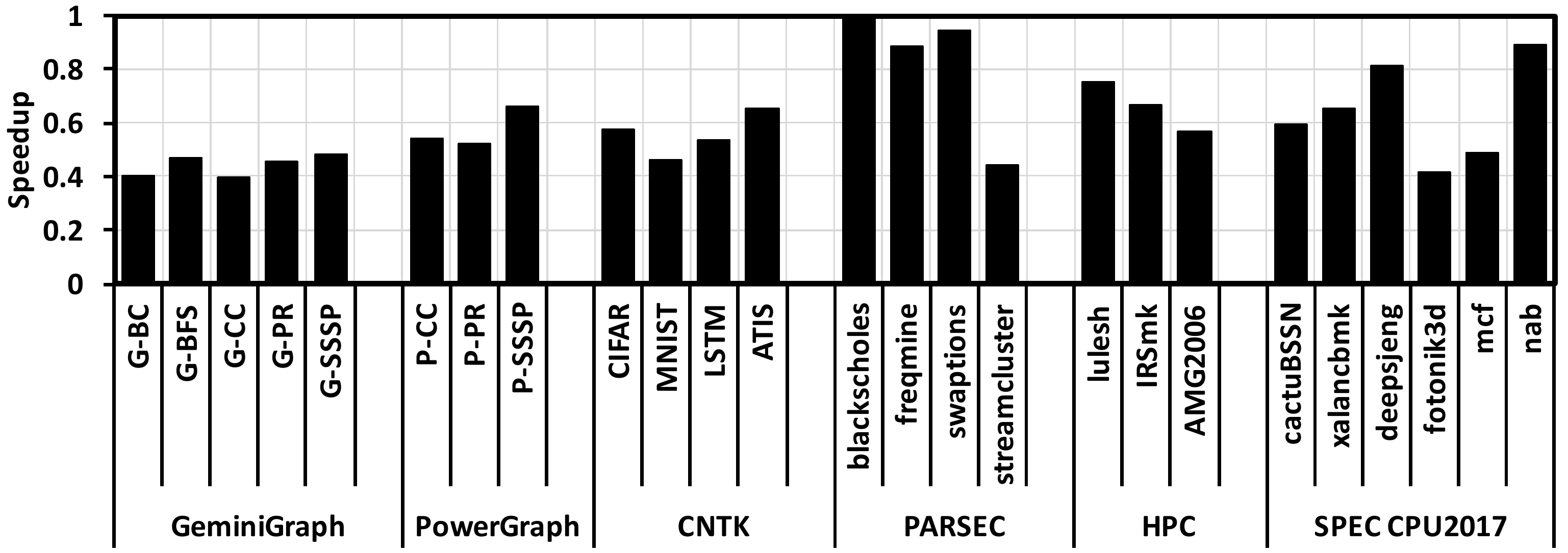}
        \caption{Co-run with Stream}
        \label{fig:corun_bandit}
    \end{subfigure}
    \caption{Speedup for applications co-running with mini-benchmarks.}
    \label{fig:corun_minibenchmarks}
    \vspace{-5mm}
\end{figure}

\begin{figure*}[t]
    \centering
    \begin{subfigure}[b]{0.23\textwidth}
        \includegraphics[width=\textwidth]{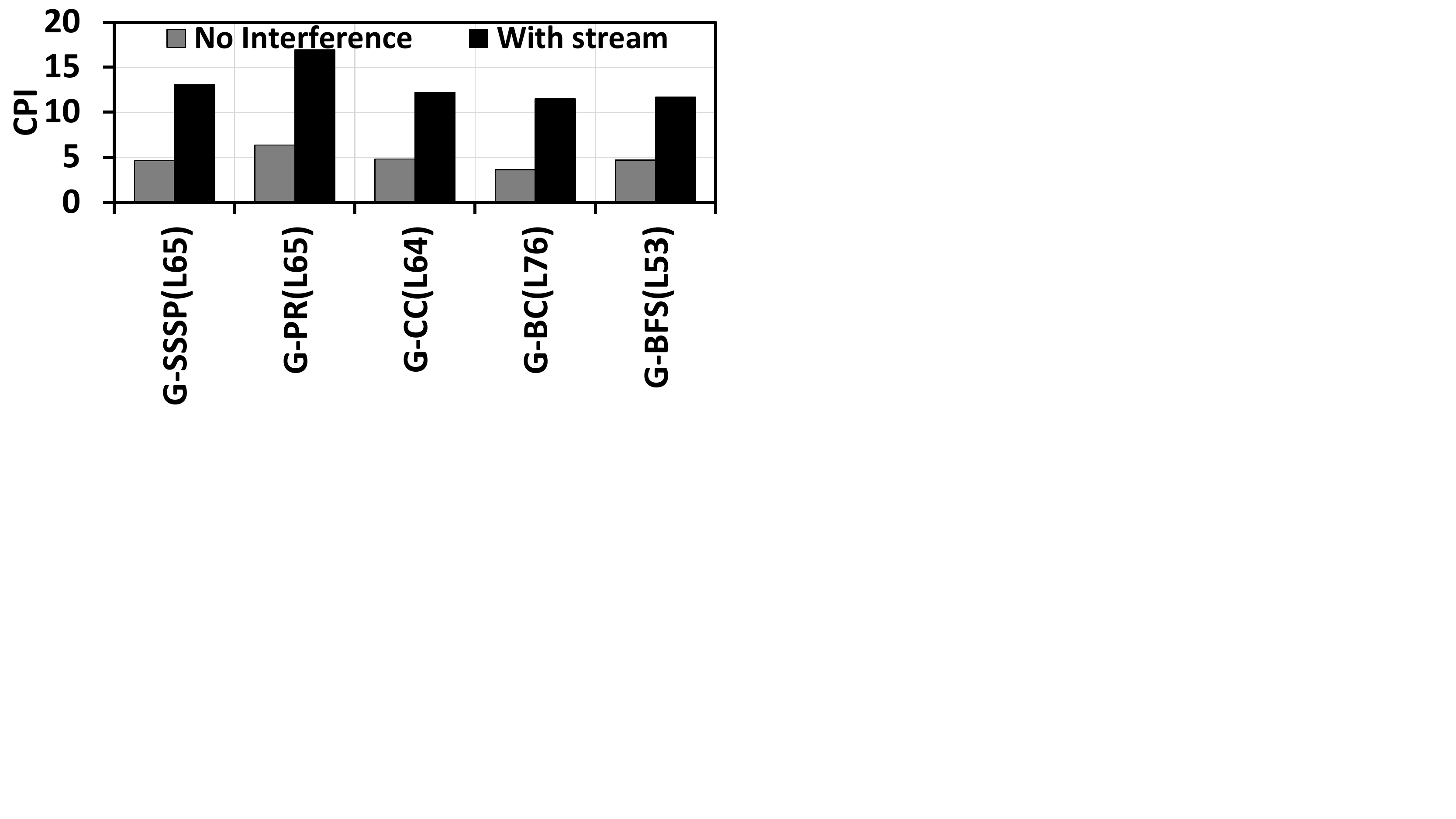}
        \caption{CPI}
        \label{fig:cpi_stream}
    \end{subfigure}  
    \begin{subfigure}[b]{0.23\textwidth}
        \includegraphics[width=\textwidth]{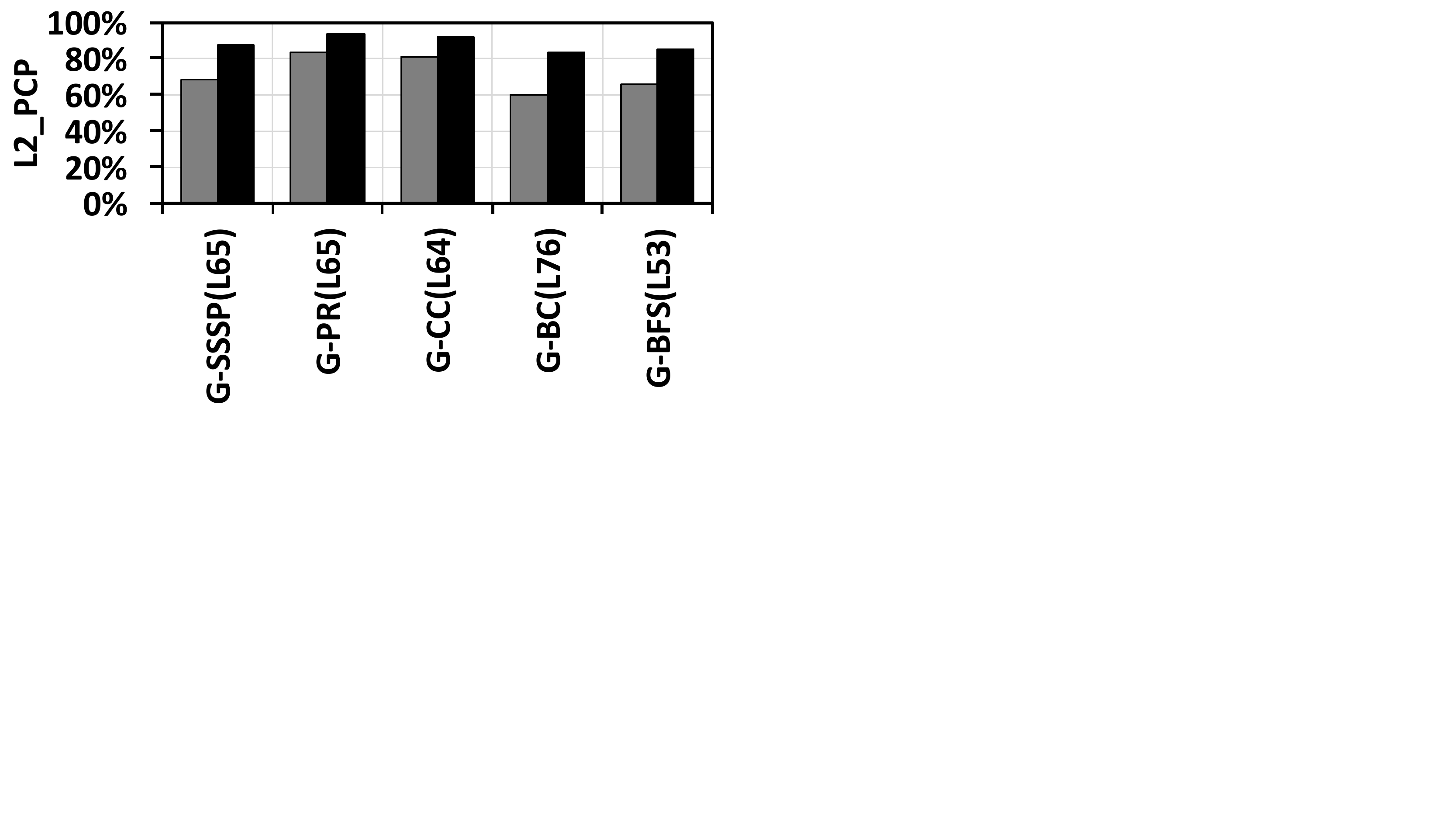}
        \caption{L2\_PCP}
        \label{fig:l2_stream}
    \end{subfigure}
    \begin{subfigure}[b]{0.23\textwidth}
        \includegraphics[width=\textwidth]{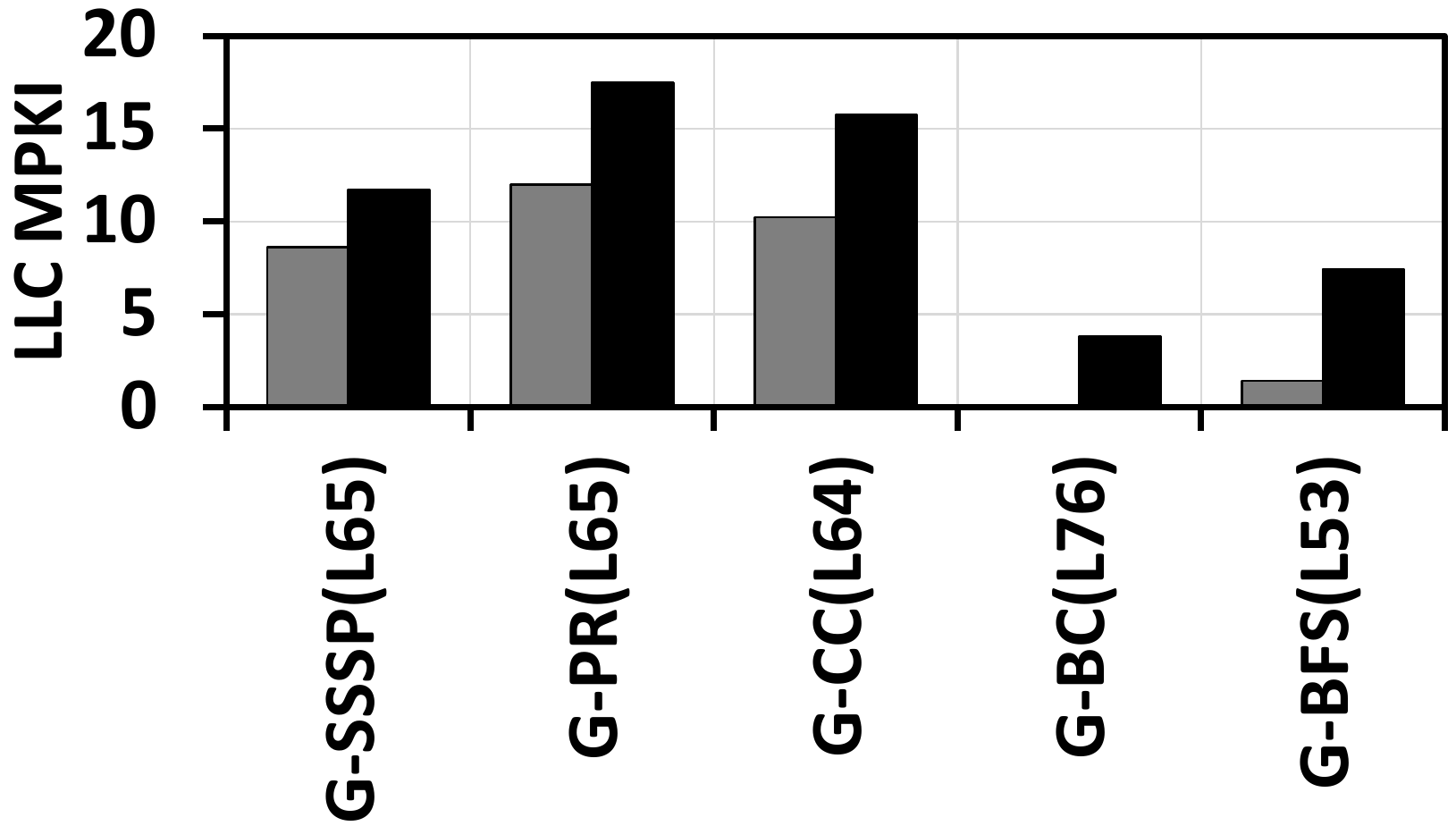}
        \caption{LLC MPKI}
        \label{fig:llcm_stream}
    \end{subfigure}
    \begin{subfigure}[b]{0.23\textwidth}
        \includegraphics[width=\textwidth]{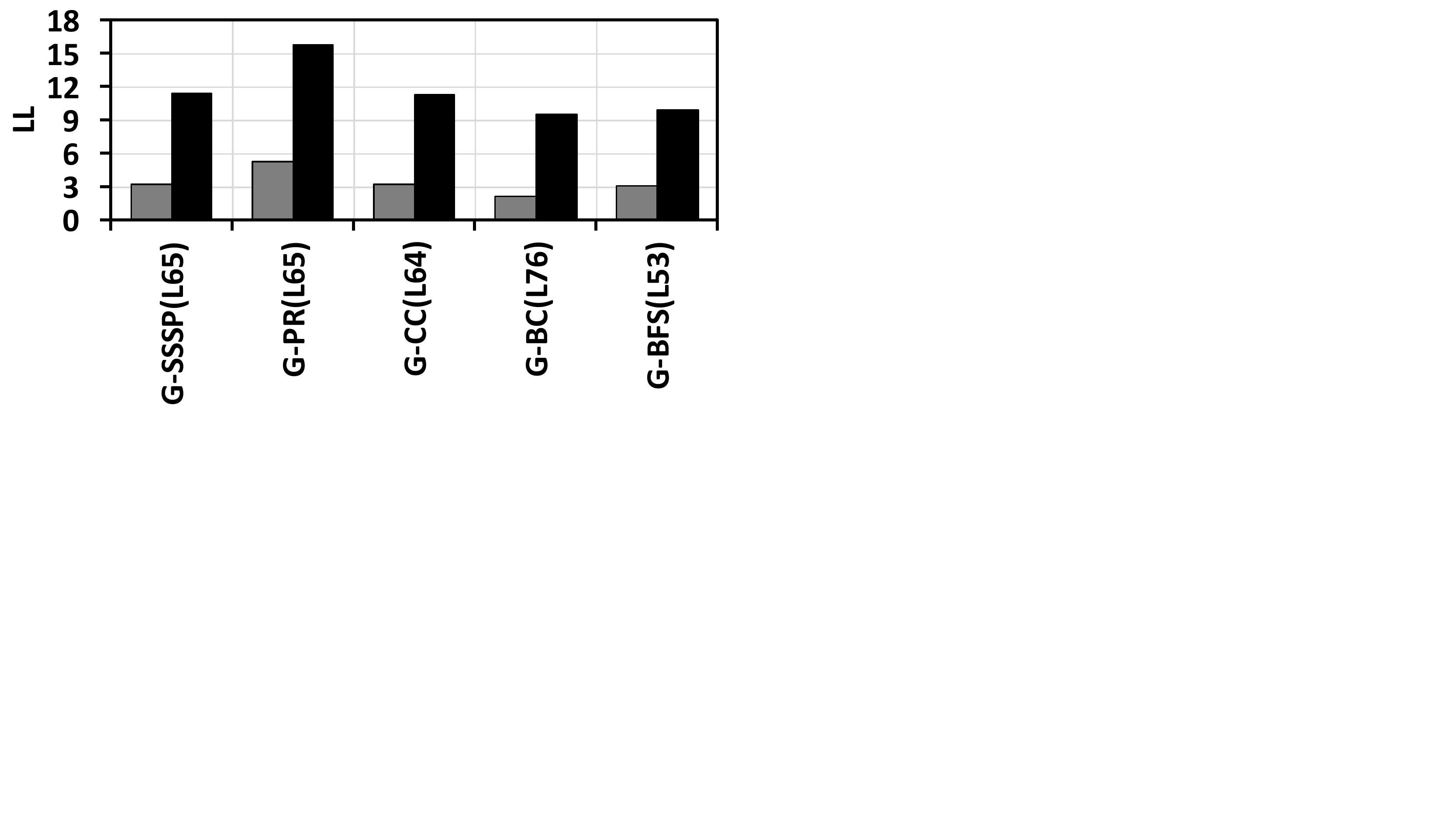}
        \caption{LL}
        \label{fig:ll_stream}
    \end{subfigure}
    \caption{Performance metrics for Gemini Graph applications co-running with Stream.}
    \label{fig:metric_stream}
    \vspace{-5mm}
\end{figure*}

\subsection{Performance Metrics for Analyzing Problematic Co-running Pairs}
To analyze the root-cause of performance degradation in the multi-tenant environment, we use the Intel Vtune profile to collect memory related hardware event statistics and map such results to the contentious code region in the application. The metrics of profiled hardware events are summarized as follows: 
\begin{itemize}
\item \texttt{CPI}: cycles per instruction.
\item \texttt{L2 misses}: the number of L2 cache misses. 
\item \texttt{L2\_PCP}: L2 Pending Cycle Percent, which is the ratio of all CPU cycles pending on L2 cache miss. 
\item \texttt{LLC MPKI}: misses of LLC per 1000 instructions.
\item \texttt{LL}: average latency of load from LLC or memory, roughly equal to average latency per L2 misses. 
\end{itemize}
Here, L2 pending cycles can be split into two parts: cycles spent on LLC and cycles spent on memory access. Both LLC and memory subsystem are the shared resources in our co-running configuration. Therefore, we use \texttt{LL} to quantify the average latency happened at shared resource, which is defined as:
$$\texttt{LL}= \frac{\texttt{CPI} * \texttt{L2\_PCP}}{\texttt{L2 misses count per instruction}}$$

In our co-running setup, two applications do not share physical cores, and each physical core owns private L1/L2 cache. Hence, \texttt{L2 misses count per instruction} is considered to be a fixed value for each application with a unvarying core count, regardless of the interference caused by task co-execution. Consequently, \texttt{LL} is driven by $\texttt{CPI}$ and $\texttt{L2\_PCP}$.

\subsection{Co-running with Mini-benchmarks}

We choose two mini-benchmarks consuming high bandwidth: Stream and Bandit, introduced in section \ref{sec:exp}. To evaluate each benchmark's sensitivity to different memory access patterns, we co-run all of the 25 applications with these mini-benchmarks. Here, mini-benchmark is deployed as a background application, which takes 4 physical cores.

Figure~\ref{fig:corun_stream} and~\ref{fig:corun_bandit} show the normalized execution time for each application when co-running with Bandit and Stream, respectively. Compared to co-running with Stream, applications co-running with Bandit have a relatively smaller slowdown ranging between $0.77\times$ to $1.0\times$. Among all applications, GeminiGraph applications' execution time gets affected the most, as the average slowdown is $0.82\times$. Even though PowerGraph applications analyze the same graph input, the interference only slows them down by an average of $0.93\times$. 
Other than the graph applications, the execution time of streamcluster and fotonik3d increase by 21\% and 27\%, respectively. Based on the Intel PCM's measurement, Bandit consumes memory bandwidth at a rate of 18GB/s, when it is running alone with 4 thread slots provided. Since Bandit doesn't benefit from cache and hardware prefetchers, it only stresses the bandwidth resource in the co-executing environments. Our experiment shows that the bandwidth contention at such level can not impact emerging applications' performance too much.

\begin{figure*}[t]
    \centering
    \begin{subfigure}[b]{0.23\textwidth}
        \includegraphics[width=\textwidth]{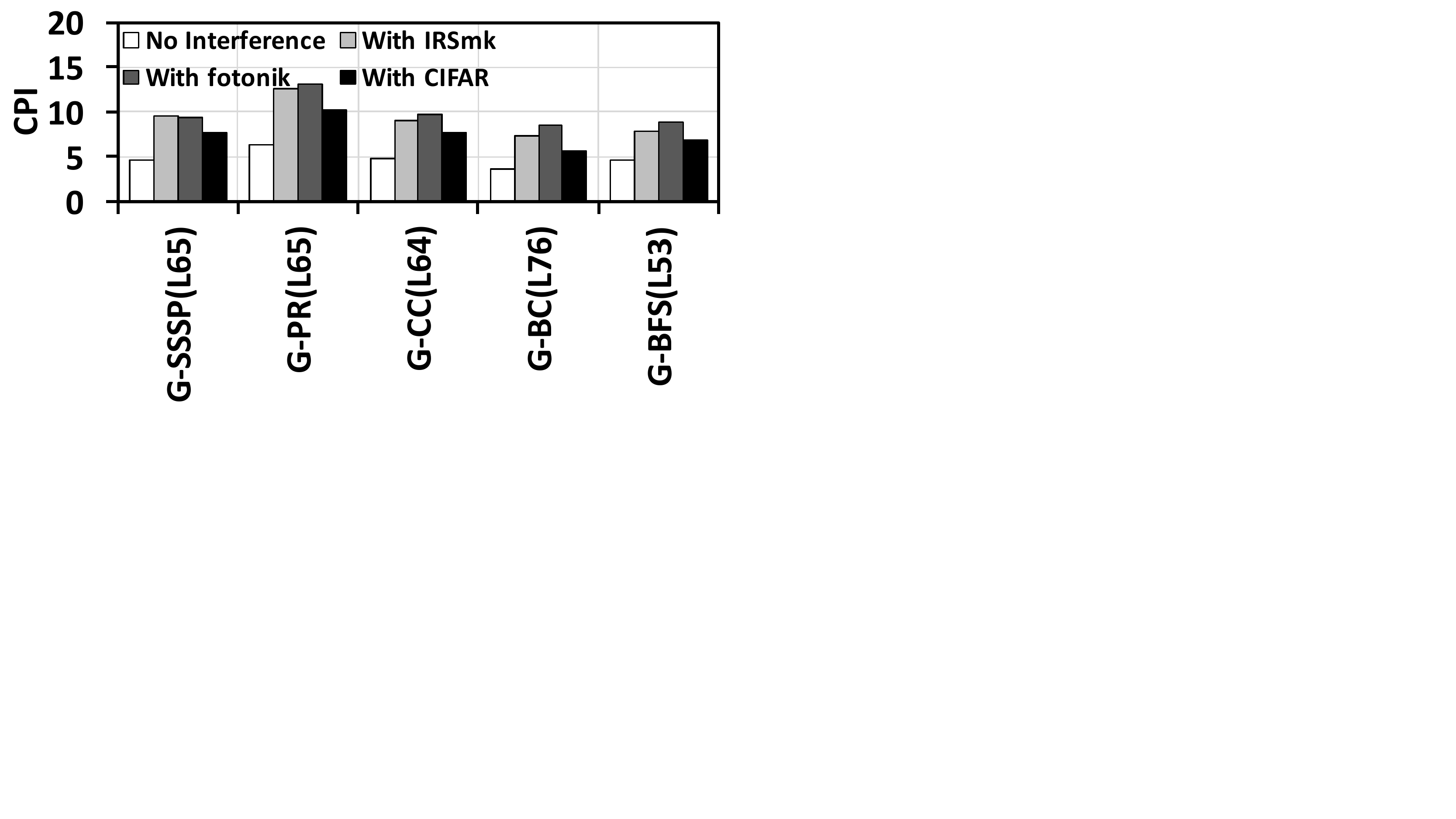}
        \caption{CPI}
        \label{fig:cpi_other}
    \end{subfigure}  
    \begin{subfigure}[b]{0.23\textwidth}
        \includegraphics[width=\textwidth]{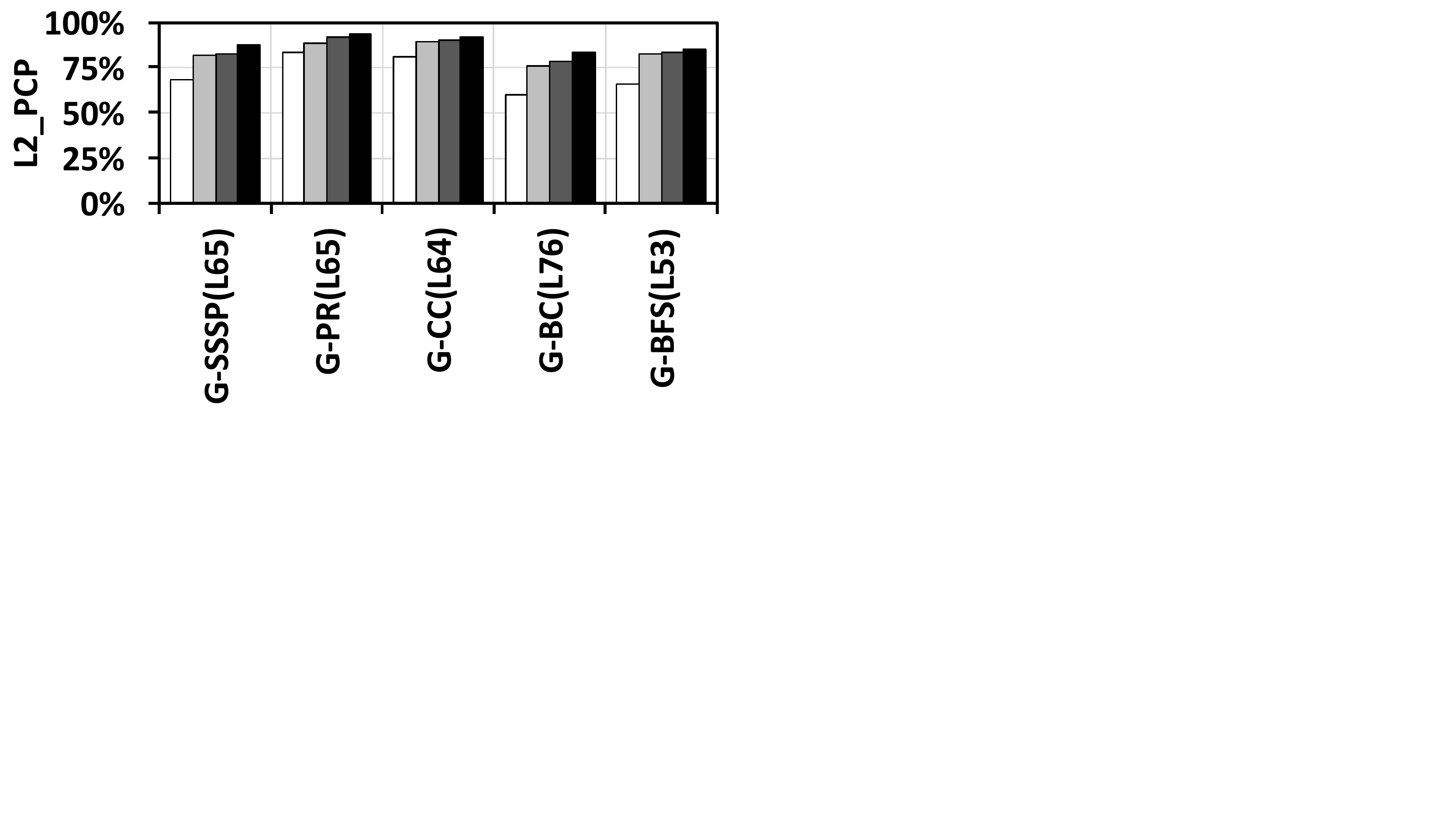}
        \caption{L2\_PCP}
        \label{fig:l2_other}
    \end{subfigure}
    \begin{subfigure}[b]{0.23\textwidth}
        \includegraphics[width=\textwidth]{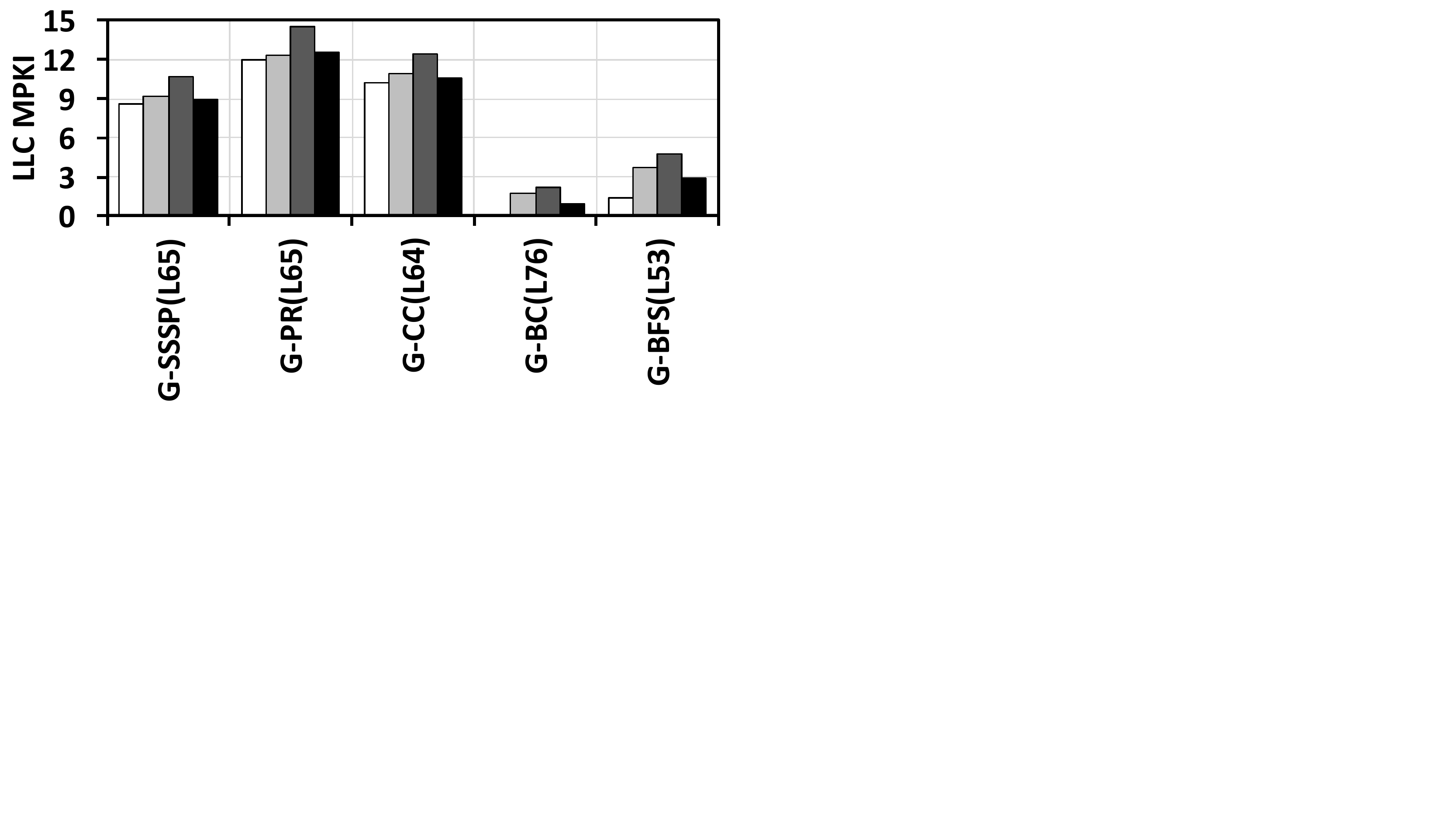}
        \caption{LLC MPKI}
        \label{fig:llcm_other}
    \end{subfigure}
    \begin{subfigure}[b]{0.23\textwidth}
        \includegraphics[width=\textwidth]{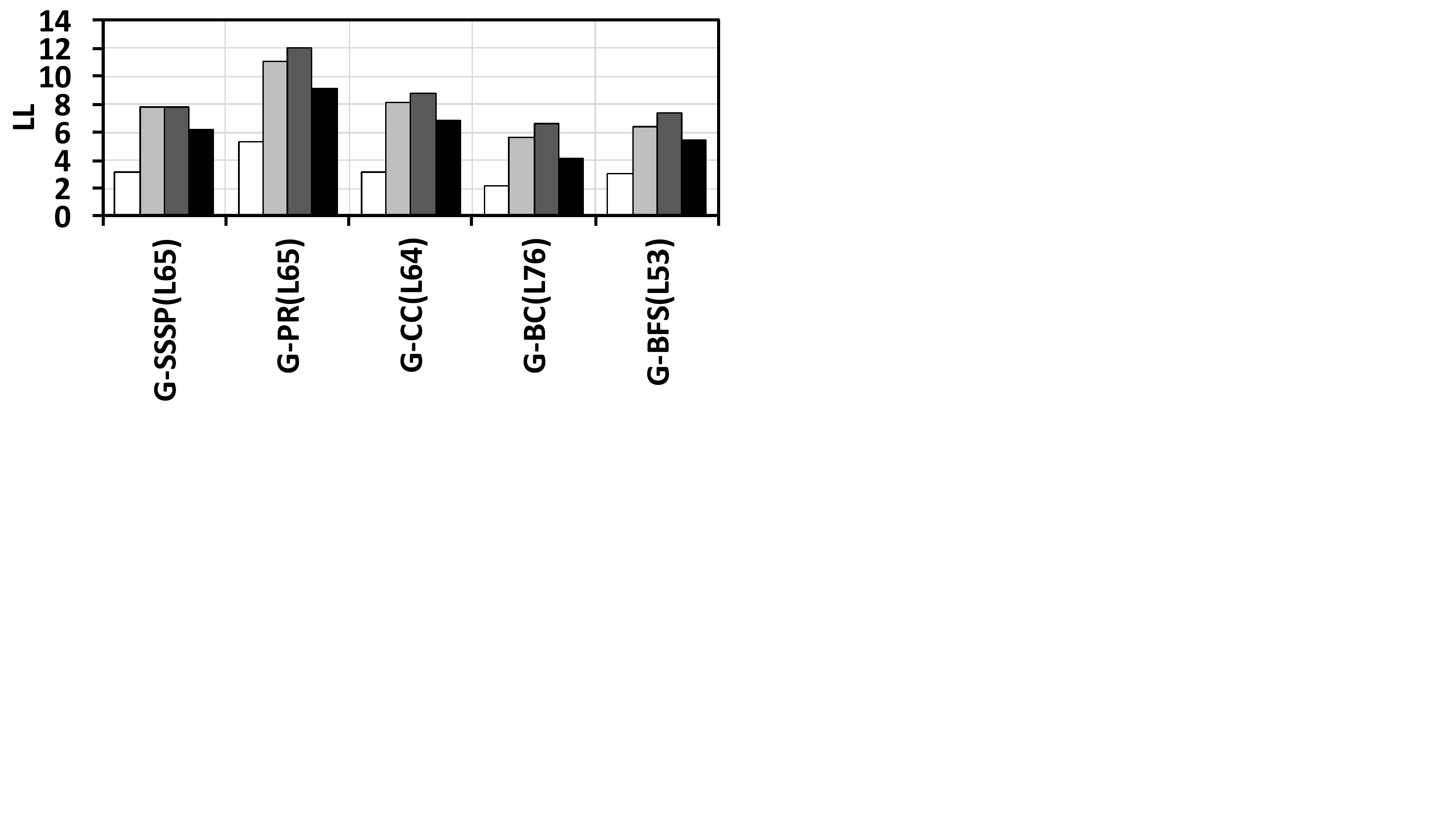}
        \caption{LLC MPKI}
        \label{fig:ll_other}
    \end{subfigure}
    \caption{Performance metrics for Gemini Graph applications co-running with offender applications.}
    \label{fig:metric_gemini_otherapps}
    \vspace{-5mm}
\end{figure*}

When applications co-locate with Stream, they suffer more in terms of performance, as the average slowdown of 25 application drops to $0.61\times$.
The average runtime of GeminiGraph applications' and PowerGraph applications' execution time is increased to $208\%$. Only few applications avoid such tremendous performance degradation: blackscholes, freqmine, swaptions, deepsjeng and nab. These benchmarks consume very little bandwidth resource, as we previously showed in Figure \ref{fig:bandwidth_eachapp}. 
The highest practical memory bandwidth of our system is approximately 28 GB/s. Running a 4-thread Steam benchmark alone can produce 24.5 GB/s bandwidth consumption. Such high bandwidth consumption is due to the regular memory access pattern that can be benefited by hardware prefetchers a lot. Other than the memory bandwidth, co-executing applications also share the LLC, which can be severely impacted by the memory intensive applications like Stream as well. Compared to other applications, We find that data analytical applications on GeminiGraph have the worst performance degradation co-running with Stream. 

Due to the space limit, Figure \ref{fig:metric_stream} only demonstrates 5 GeminiGraph applications' comparisons of co-running with Stream and running alone without interference in terms of \texttt{LLC MPKI}, \texttt{CPI}, \texttt{L2\_PCP}, and \texttt{LL}.
As shown in Figure~\ref{fig:llcm_stream}, the \texttt{LLC MPKI} of each application also increases by $2.6\times$ due to the LLC contention. Both Stream and graph applications benefit from the LLC. However, 
two applications compete for the limited space of last level cache, causing the increment of LLC misses for graph applications. The combination of the interference effect from the LLC and memory bandwidth results in high \texttt{LC\_PCP}, which means most of the execution cycles are spent on waiting for the data loaded from LLC or memory. Especially for G-PR, \texttt{LC\_PCP} reaches $93\%$ of total cycles. Such memory bottleneck worsens the overall \texttt{CPI} of each graph application. Figure~\ref{fig:cpi_stream} shows that every application's \texttt{CPI} increases more than $2\times$, which is highly correlated to the overall runtime elongation. Consequently, \texttt{LL} driven by \texttt{CPI} and \texttt{L2\_PCP} also gets a more than $2\times$ increase for each application (shown in Figure~\ref{fig:l2_stream}). Compared to other graph applications in GeminiGraph, G-BFS and G-BC are less memory-intensive, as these applications only iterate and update each vertex once. Such lightweight memory access can be well serviced by the LLC cache. However, when they are sharing the LLC with Stream, \texttt{LLC MPKI} of both application dramatically increases, as shown in Figure~\ref{fig:llcm_stream}.

The co-running experiments with two mini-benchmarks help to verify the conclusion in the previous section: applications with the following two properties :
\begin{itemize}
\item Large amount of memory accesses with regular pattern 
\item Consuming large portion of bandwidth
\end{itemize}
will easily degrade performance of memory intensive applications. Applications with the such properties can also impact graph benchmarks a lot, we will analysis it in the next subsection.

\begin{table}[]
\centering
\scriptsize
\tabcolsep=0.11cm
\caption{Profiling results of P-PR and fotonik3d.}
\begin{tabular}{|c|c|c|c|c|c|c|}
\hline
Metric & \begin{tabular}[c]{@{}c@{}}Foreground\\ Application\end{tabular} & \begin{tabular}[c]{@{}c@{}}No \\ Interference\end{tabular} & \begin{tabular}[c]{@{}c@{}}With \\ IRSmk\end{tabular} & \begin{tabular}[c]{@{}c@{}}With \\ CIFAR\end{tabular} & \begin{tabular}[c]{@{}c@{}}With \\ fotonik3d\end{tabular} & \begin{tabular}[c]{@{}c@{}}With \\ G-SSSP\end{tabular} \\ \hline
\multirow{2}{*}{CPI} & \begin{tabular}[c]{@{}c@{}}P-PR\\ (gather)\end{tabular} & 2.3 & 3.7 & 3.5 & 4.3 & - \\ \cline{2-7} 
 & \begin{tabular}[c]{@{}c@{}}fotonik3d\\ (UUS)\end{tabular} & 2.0 & 3.6 & 3.2 & - & 1.8 \\ \hline
\multirow{2}{*}{\begin{tabular}[c]{@{}c@{}}LLC \\ MPKI\end{tabular}} & \begin{tabular}[c]{@{}c@{}}P-PR\\ (gather)\end{tabular} & 3.9 & 5.2 & 4.8 & 4.8 & - \\ \cline{2-7} 
 & \begin{tabular}[c]{@{}c@{}}fotonik3d\\ (UUS)\end{tabular} & 20.9 & 22.1 & 21.1 & - & 21.3 \\ \hline
\multirow{2}{*}{L2\_PCP} & \begin{tabular}[c]{@{}c@{}}P-PR\\ (gather)\end{tabular} & 71\% & 80\% & 79\% & 83\% & - \\ \cline{2-7} 
 & \begin{tabular}[c]{@{}c@{}}fotonik3d\\ (UUS)\end{tabular} & 65\% & 80\% & 81\% & - & 63\% \\ \hline
\multirow{2}{*}{LL} & \begin{tabular}[c]{@{}c@{}}P-PR\\ (gather)\end{tabular} & 1.7 & 2.9 & 2.8 & 3.6 & - \\ \cline{2-7} 
 & \begin{tabular}[c]{@{}c@{}}fotonik3d\\ (UUS)\end{tabular} & 1.3 & 2.9 & 2.6 & - & 1.2 \\ \hline
\end{tabular}
\label{table:presults_pr_fotonik3d}
\vspace{-5mm}
\end{table}

\subsection{GeminiGraph Applications}
\label{sec:cs_gemini}
In section \ref{sec:interference}, we demonstrate graph applications suffer from co-running with fotonik3d, IRSmk and CIFAR applications. Figure \ref{fig:metric_gemini_otherapps} shows \texttt{CPI}, \texttt{L2\_PCP} and \texttt{MPKI} comparison of co-running with three \textbf{Offender} applications and running alone without interference.  The \texttt{LLC MPKI} of each application increases up to $18\%$. All applications benefit from LLC. The interference on LLC caused by these three \textbf{Offender} applications is not as severe as Stream. However, high \texttt{L2\_PCP} of all co-running pairs means LLC and memory subsystem are the bottleneck. 
Correspondingly, \texttt{LL} also increases more than $100\%$. 

Similar to Stream, Fotonik3d and IRSmk have a very regular memory access pattern. However, they are not as harmful as Stream, as purely regular memory access pattern is impossible to achieve in the real world applications. Since CIFAR's memory access pattern is less regular than fotonik3d and IRSmk, CIFAR's impact on graph applications are much less than IRSmk and fotonik3d. As shown in Figure \ref{fig:corun_heatmap}, only GeminiGraph applications co-running with CIFAR have a normalized execution time below 1.5, except for G-CC.    

\lstset { %
    language=C++,
    backgroundcolor=\color{white}, 
    breaklines=true,
    basicstyle=\scriptsize\ttfamily,
    numbers=left,
    stepnumber=1,
    firstnumber=63,
    xleftmargin=2em,
}
\begin{figure}[h]
\centering
\scriptsize
\begin{lstlisting}
[&](VertexId dst, VertexAdjList<Empty> incoming_adj) {
  double sum = 0;
  for (AdjUnit<Empty> * ptr=incoming_adj.begin; ptr!=incoming_adj.end; ptr++) {
    VertexId src = ptr->neighbour;
    sum += curr[src];
  }
  graph->emit(dst, sum);
}
\end{lstlisting}
\caption{Code for Pagerank Gemini (G-PR), line number from $63$ to $70$ for pagerank.c}
\label{fig:code_pagerank_gemini}
\vspace{-7mm}
\end{figure}

\begin{figure}[h]
\centering
\scriptsize
\begin{lstlisting}
double gather(icontext_type & context, const vertex_type& vertex, edge_type& edge) const{
	// printf("%ld", edge.source().num_out.edges());
	return (edge.source().data()/edge.source().num_out_edges());
}
\end{lstlisting}
\caption{Code for Pagerank Powergraph (P-PR), line number from $63$ to $66$ for pagerank.c}
\label{fig:code_pagerank_powergraph}
\vspace{-5mm}
\end{figure}

\subsection{PowerGraph Applications}
\label{sec:cs_power}
P-PR is the only graph application in PowerGraph that suffers from co-running with fotonik3d, CIFAR and IRSmk. For P-PR, function \texttt{gather} takes most of the CPU execution cycles. The code of \texttt{gather} is demonstrated in Figure \ref{fig:code_pagerank_powergraph}. The \texttt{gather} function is responsible for loading all the value of connected edges. Such massive data loading phase in graph applications (implemented in gather-apply-scatter execution model) can be heavily impacted, when co-running with fotonik3d, CIFAR and IRSmk. Identifying the contentious code region in the graph application can benefit system designers to carry out more robust graph computation model. Meanwhile, it can help compiler designer to relief the interference impact at the source code level.

Similar to GeminiGraph applications, P-PR also benefits from the LLC. As shown in Table \ref{table:presults_pr_fotonik3d}, \texttt{LLC MPKI} increases nearly $30\%$. Three \textbf{Offender} applications cause intensive \texttt{LLC} contention. Due to the high \texttt{L2\_PCP} (up to $83\%$) and high \texttt{LL}, P-PR is bounded by LLC and memory bandwidth. The memory burden of PowerGraph is less than GeminiGraph, and the performance of PowerGraph is worse than GeminiGraph. As a result, the interference impact brought by three \textbf{Offender} applications (fotonik3d, IRSmk, CIFAR) on PowerGraph is much less than that on GeminiGraph.

\subsection{Fotonik3d}
Even though fotonik3d offends other co-running applications severely (e.g. graph applications), it is impacted by other \textbf{Offender} applications like CIFAR and IRSmk. 
In Table \ref{table:presults_pr_fotonik3d}, Fotonik3d's profiling results are compared between co-runs and sole-run.
Fotonik3d's \texttt{LLC MPKI} is more than $20$ when running alone without interference. It doesn't change too much after co-running with the other two \textbf{Offender} applications. Therefore, \texttt{LLC} contention is not the root-cause for the performance loss. However, \texttt{L2\_PCP} increases from $65\%$ to nearly $80\%$, which shows that memory bandwidth becomes the bottleneck. Memory bandwidth bottleneck causes \texttt{LL} increase more than $1\times$. Both fotonik3d and IRSmk have a very regular memory access pattern and consume large portion of available bandwidth. When they are formed as a co-execution pair, fotonik3d suffers more than IRSmk. 
Compared to consolidating with CIFAR, fotonik3d losses more performance, when it shares the hardware resource with IRSmk. Such performance degradation is not observed in the co-execution with G-SSSP.

%% file: conclusion.tex
In this paper, we systematically study the interference when multiple applications co-run on the same machine. We empirically investigate 25 modern programs from various application domains, such as data mining, machine learning, standard benchmarks, HPC, and real-world parallel applications, among others. We conduct 625 pairs of co-running settings to obtain deep insights of understanding execution interference and identify its root causes in both hardware and software. Our findings help understand the characteristics (e.g., harmony, victim-offender, or both-victim) of individual applications to execution interference. Moreover, the root cause analysis associates the interference sensitivity with source code regions. The insights provided by this paper will benefit both hardware and software design for throughput-oriented computing systems. Due to the page limit, some interesting findings are omitted. Hence, a repository that contains all the experiment results will be provided after acceptance of the paper.